\documentclass{aa}

\usepackage{graphicx} 
\usepackage{amsmath}
\usepackage{upgreek}
\usepackage{float}
\usepackage{hyperref}
\usepackage{amsmath}
\usepackage{xspace}
\usepackage{multirow}
\usepackage{xcolor}
\usepackage{natbib}
\usepackage{enumitem}

\bibpunct{(}{)}{;}{a}{}{,}

\newcommand{\OD}{$\log(1+\delta_{\rm gal})$\xspace}
\newcommand{\pz}{photo-$z$\xspace}
\newcommand{\sz}{spec-$z$\xspace}
\newcommand{\gz}{grism-$z$\xspace}

\def\orcid#1{\href{https://orcid.org/#1}{\includegraphics[scale=0.018]{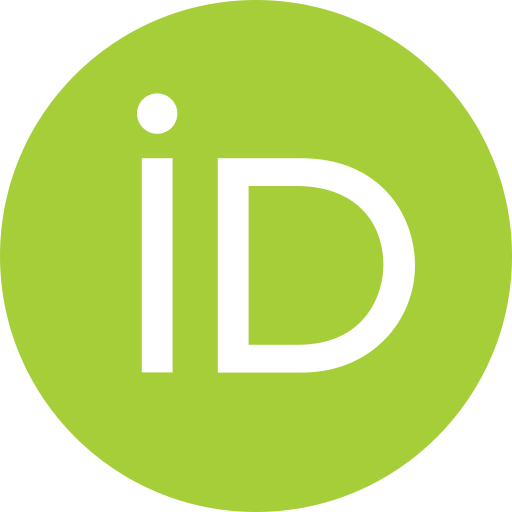}}}

\graphicspath{{./}{Figures/}}

\begin{document}

\title{The \emph{HST}-Hyperion survey: Environmental imprints on the stellar mass function at $z\sim 2.5$}
\titlerunning{\emph{HST}-Hyperion GSMF}
\authorrunning{Sikorski et al.}

\author{
        Derek Sikorski\orcid{0000-0001-5796-2807}\inst{\ref{IfA}}\thanks{Email: dsikors@hawaii.edu}
        \and Ben Forrest\orcid{0000-0001-6003-0541}\inst{\ref{UCD}}
        \and Brian C. Lemaux\orcid{0000-0002-1428-7036}\inst{\ref{UCD}\and\ref{Gemini}}
        \and Lu Shen\orcid{0000-0001-9495-7759}\inst{\ref{A&M1}\and\ref{A&M2}}
        \and Finn Giddings\orcid{0009-0003-2158-1246}\inst{\ref{IfA}}
        \and Roy Gal\orcid{0000-0001-8255-6560}\inst{\ref{IfA}}
        \and Olga Cucciati\orcid{0000-0002-9336-7551}\inst{\ref{INAFB}}
        \and Emmet Golden-Marx\orcid{0000-0001-5160-6713}\inst{\ref{INAFP}}
        \and Weida Hu\inst{\ref{A&M1}\and\ref{A&M2}}
        \and Denise Hung\orcid{0000-0001-7523-140X}\inst{\ref{Gemini}}
        \and Lori Lubin\orcid{0000-0003-2119-8151}\inst{\ref{UCD}}
        \and Kaila Ronayne\orcid{0000-0001-5749-5452}\inst{\ref{A&M1}\and\ref{A&M2}}
        \and Ekta Shah\orcid{0000-0001-7811-9042}\inst{\ref{UCD}}
        \and Sandro Bardelli\orcid{0000-0002-8900-0298}\inst{\ref{INAFB}}
        \and Devontae C. Baxter\orcid{0000-0002-8209-2783}\inst{\ref{UCSD}}
        \and Gayathri Gururajan\orcid{0000-0002-7472-7697}\inst{\ref{SISSA}}
        \and Laurence Tresse\orcid{0000-0001-8776-0958}\inst{\ref{Aix}}
        \and Gianni Zamorani\orcid{0000-0002-2318-301X}\inst{\ref{INAFB}}
        \and Joel Diamond\orcid{0009-0003-1792-6539}\inst{\ref{Charleston}}
        \and Lucia Guaita\orcid{0000-0002-4902-0075}\inst{\ref{Chile}}
        \and Nimish Hathi\orcid{0000-0001-6145-5090}\inst{\ref{STSI}}
        \and Elena Zucca\orcid{0000-0002-5845-8132}\inst{\ref{INAFB}}
}

   \institute{
    Institute for Astronomy, University of Hawai‘i, 2680 Woodlawn Drive, Honolulu, HI 96822, USA \label{IfA}
    \and Department of Physics and Astronomy, University of California, Davis, One Shields Avenue, Davis, CA 95616, USA\label{UCD}    
    \and Gemini Observatory, NSF NOIRLab, 670 N. A’ohoku Place, Hilo, HI, 96720, USA \label{Gemini}
    \and Department of Physics and Astronomy, Texas A\&M University, College Station, TX, 77843 USA \label{A&M1}
    \and George P. and Cynthia Woods Mitchell Institute for Fundamental Physics and Astronomy, Texas A\&M University, College Station, TX, 77843-4242 USA \label{A&M2}
    \and  INAF Osservatorio di Astrofisica e Scienza dello Spazio di Bologna, Via Piero Gobetti 93/3, 40129 Bologna, Italy \label{INAFB}
    \and INAF - Osservatorio astronomico di Padova, Vicolo Osservatorio 5, 35122 Padova, Italy \label{INAFP}
    \and Department of Astronomy \& Astrophysics, University of California, San Diego, 9500 Gilman Dr, La Jolla, CA 92093, USA \label{UCSD}
    \and Scuola Internazionale Superiore Studi Avanzati (SISSA), Physics Area, Via Bonomea 265, 34136 Trieste, Italy \label{SISSA}
    \and Aix-Marseille Université, CNRS, CNES, LAM, Marseille, France \label{Aix}
    \and Department of Physics \& Astronomy, College of Charleston, 58 Coming Street, Charleston, SC 29424, USA \label{Charleston}
    \and Universidad Andres Bello, Facultad de Ciencias Exactas, Departamento de Fisica y Astronomia, Instituto de Astrofisica, Fernandez Concha 700, Las Condes, Santiago RM, Chile \label{Chile}
    \and Space Telescope Science Institute, 3700 San Martin Drive, Baltimore, MD 21218, USA. \label{STSI}
    }

\abstract
    {
        Not all galaxies at cosmic noon ($2 \lesssim z \lesssim 3$) evolve in the same way. Particularly, it remains unclear how and to what extent the local environment -- especially the extreme overdensities of protoclusters -- affects the stellar mass assembly of its constituent galaxies at high redshifts. The imprint of these early processes is encoded in the galaxy stellar-mass function (SMF); comparing SMFs across environments therefore reveals differences in evolutionary history.    
    }
   {
        We present the SMF of the Hyperion proto-supercluster at $z\sim 2.5$,  one of the largest and most massive protostructures in the early universe. This dataset yields the most statistically robust SMF of a single protostructure at $z\gtrsim 2$. By comparing the SMF of the overdense peaks within Hyperion to the coeval field, we begin to answer the question of how early, and how strongly,  a dense environment tilts the balance in favor of massive galaxies
    } 
   {
        Given that Hyperion resides in the field of the Cosmic Evolution Survey (COSMOS), we combined the extensive COSMOS2020 photometric catalog with ground-based spectroscopy and new grism spectroscopy from the Hubble Space Telescope (\emph{HST}). The structure of Hyperion is defined based on a three-dimensional overdensity map, allowing us to place galaxies into  (i) the highly overdense {peaks} of Hyperion, (ii) the less-overdense {outskirts} of Hyperion, or (iii) a coeval field. We performed 100 Monte Carlo realizations of the data to propagate redshift and stellar mass uncertainties, refitting galaxy properties in each realization. After constructing SMFs for the outskirts and peaks of Hyperion, we normalized them to that of the field to highlight differences in the underlying shape of the SMFs.
   }
   {
        The overdense peaks of Hyperion host a striking excess of massive galaxies relative to the field: the number densities of $\log_{10}(M_*/M_\odot)\sim 11$ galaxies are $\sim 10\times$ higher than the coeval field, whereas $\log_{10}(M_*/M_\odot)\sim 9.5$ galaxies are enhanced by only $\sim 3.5\times$. On the other hand, both the SMF of the outskirts of Hyperion and the SMF of Hyperion as a whole mirror the overall shape of the coeval field.
    }
   {
        Environmental effects that govern stellar mass growth are already well established by $z\sim 2.5$. The densest regions of Hyperion  host   galaxies that have already experienced accelerated stellar mass growth. Furthermore, this impact is largely masked in the total SMF of Hyperion, highlighting the necessity of deep spectroscopic surveys when uncovering environmental trends at high redshifts. These findings imply that high-redshift protostructures begin sculpting the high-mass end of the SMF well before the epoch when local clusters experience widespread quenching, and may provide the appropriate laboratories for producing the elevated star formation observed at cosmic noon.
   }

    \keywords{
                    Galaxies: clusters: individual: Hyperion —
                    Galaxies: evolution —
                    Galaxies: mass function
            }

\maketitle
\nolinenumbers

\section{Introduction}

    It has long been understood that different galaxies can experience dramatically different evolutionary pathways, thus resulting in distinct physical properties. This is particularly evident at low and intermediate redshifts ($z \lesssim 1.5$) where galaxies residing in dense cores of clusters at low redshifts tend to display higher stellar masses \citep[e.g.,][]{Kauffmann+04, Baldry+06, VanderBurg+20}, lower star formation rates \citep[SFRs; e.g.,][]{Gomez+03, vonderLinden+10, Muzzin+12, Tomczak+19, Old+20}, and redder colors \citep[e.g.,][]{Peng+10, Lemaux+12, Nantais+16, Lemaux+19, VanderBurg+20} compared to their counterparts in the coeval field. Under the assumption that these two apparently different populations of galaxies formed with similar initial properties, this bimodality implies the existence of environmentally dependent processes that have profound effects on the way galaxies evolve. However, the exact nature of the dominant processes responsible for this differential evolution and the time period in which they act  remain elusive.

    By cosmic noon ($2 \lesssim z \lesssim 3$), the picture seemingly flips. Several studies now suggest that overdense environments are conducive to {higher} star formation rate densities (SFRDs) than the field \citep[e.g.,][]{Lemaux+22, Staab+24}. This is a stark contrast to the local universe, and suggests a more complex redshift-dependence of the environmentally dependent processes driving galaxy evolution. Taken together with local trends, it seems as though the same environments that eventually suppress stellar mass growth in the local universe may in fact accelerate it in the early universe.

    Given that the key mechanisms spurring galaxy evolution likely take place over prolonged timescales, catching them in the act for any given galaxy is difficult. However, their integrated impact is encoded in the {stellar mass function} (SMF), i.e., the number density of galaxies as a function of stellar mass. In the local universe, various studies focused on the SMF have contributed to a better understanding of the processes driving the observed bimodality of galaxies. In particular, \citet{Peng+10} developed an empirical model in which {environmental quenching} truncates star formation irrespective of mass, while {mass quenching} limits growth in the most massive systems regardless of environment. At intermediate redshifts, studies such as \citet{Tomczak+17} find the SMF depends strongly on environment: overdense environments host an excess of higher stellar mass compared to lower stellar mass galaxies, relative to the field.

    Owing to the sparseness of data, studies of SMFs in overdense environments at $z\gtrsim 2.0$ have been few and far between, and show a variety of results. Using narrowband imaging of H$\alpha$-emitters, studies suggest an enhancement of higher-mass galaxies in protoclusters at $z\sim 2.2$ and $z\sim 2.5$ \citep{Shimakawa+18b, Shimakawa+18a}. On the other hand, \citet{Edward+24} find that a composite SMF built from several $2.0<z<2.5$ protoclusters shows little deviation from the field. At even higher redshifts, \citet{Forrest+24} find the enhancement of high-mass galaxies is also present in the Elent\'ari proto-supercluster at $z\sim 3.3$, particularly in its most overdense regions.

    To this end, we present an analysis of the SMF within the Hyperion proto-supercluster at $z\sim 2.5$, the largest and most massive structure in the field of the Cosmic Evolution Survey \citep[COSMOS;][]{Scoville+07}. By leveraging a combination of the extensive COSMOS2020 photometric catalog \citep{Weaver+22}, a variety of ground-based spectroscopic surveys, and new grism spectroscopy using the Hubble Space Telescope (\emph{HST}) as part of the \emph{HST}-Hyperion Survey \citep{Forrest+25, Giddings+26}, we constructed what is, to our knowledge, the most statistically robust SMF of a single structure at $z\gtrsim2$. In Section \ref{Sect:Data} we describe the various sources of data used in this study. In Section \ref{Sect:GalEnvs} we describe the framework used for defining protostructures before constructing the SMF of Hyperion in Section \ref{Sect:SMF}. Finally, we discuss the results in Section \ref{Sect:Discussion} and conclude in Section \ref{Sect:Conclusion}. Throughout this study all magnitudes are presented in the AB system \citep{OkeGunn83, Fukugita+96}. We also use a $\Lambda$CDM cosmology with $\rm H_0 = 70\,\,km\,s^{-1}\,Mpc^{-1}$, $\Omega_\Lambda=0.73$, and $\Omega_M = 0.27$.

\section{Data}\label{Sect:Data}

    The Hyperion proto-supercluster is located within the footprint of the Cosmic Evolution Survey \citep[COSMOS;][]{Scoville+07}, an extensively studied field with abundant photometric and spectroscopic data. This work takes advantage of these datasets, along with targeted spectroscopy of Hyperion from ground-based observatories and \emph{HST}-grism observations. In this section, we summarize the various data sources used in this study and provide an updated picture of Hyperion.

    \subsection{COSMOS2020 photometry}\label{sSect:C20}

        The photometric measurements used in this work were taken from the COSMOS2020 catalogs \citep{Weaver+22}. These catalogs contain extensive data for $\sim \! 1.7$ million objects spanning $\sim \!2 \,\,\rm deg^2$, including imaging in up to 40 bandpasses and physical properties estimated from two spectral energy distribution (SED) fitting codes. For this work, we specifically use COSMOS2020 Classic Catalog v2.0 and physical parameters derived with \texttt{LePhare}\footnote{\texttt{LePhare}'s website: \href{http://www.cfht.hawaii.edu/~arnouts/lephare.html}{http://www.cfht.hawaii.edu/~arnouts/lephare.html}} \citep{Arnouts+99, Ilbert+06}.

        The COSMOS2020 photometry spans a wide wavelength range, from the far-ultraviolet (FUV) to the mid-infrared (MIR). While a more detailed summary of the instruments and bandpasses used in the catalog is provided in Table 1 of \citet{Weaver+22}, we briefly summarize the key data here. Space-based imaging includes FUV and near-ultraviolet (NUV) data from the Galaxy Evolution Explorer \citep[\emph{GALEX};][]{Zamojski+07}, optical data from the F814W band on \emph{HST}'s Advanced Camera for Surveys \citep[ACS;][]{Leauthaud+07}, and MIR data from channels 1-4 of the Infrared Array Camera (IRAC) aboard \emph{Spitzer} \citep{Ashby+13, Ashby+15, Ashby+18, Steinhardt+14}. Ground-based imaging includes NUV data from the $u$ and $u^*$ bands via the MegaCam instrument on the Canada-France-Hawai`i Telescope \citep[CFHT;][]{Laigle+16, Sawicki+19}, optical data from 7 broad-, 12 medium-, and 2 narrow-bands via Subaru/Suprime-Cam \citep{Taniguchi+07, Taniguchi+15}, optical data from the $grizy$ bands via Subaru/Hyper Suprime-Cam \citep{Aihara+19}, and near-infrared (NIR) data from the $YJHK_s$ broad-bands and $NB118$ narrow-band via the UltraVISTA survey \citep{McCracken+12, Moneti+23}. Astrometric measurements are taken from Gaia DR1 and DR2 \citep{Gaia+16, Gaia+18}.

        In this work we used flux densities for all bands where available, except for the \emph{HST}/ACS F814W band and \emph{Spitzer}/IRAC channels 3 and 4. Specifically, for a given band \texttt{XXXX}, we used the flux-density reported under \texttt{XXXX\_FLUX\_APER3} in the COSMOS2020 catalog and its associated error, which was derived using a $3"$ aperture in \texttt{SExtractor} \citep{Bertin+96}. For each source, an aperture-to-total flux correction using \texttt{total\_off3} was applied to the flux density in all bands taken from ground-based observations. Finally, a foreground attenuation correction was applied using the extinction term \texttt{EBV\_MW}, along with a multiplicative band-dependent factor, as described in \citet{Laigle+16}.

\begin{figure}
            \centering
            \includegraphics[width=\linewidth]{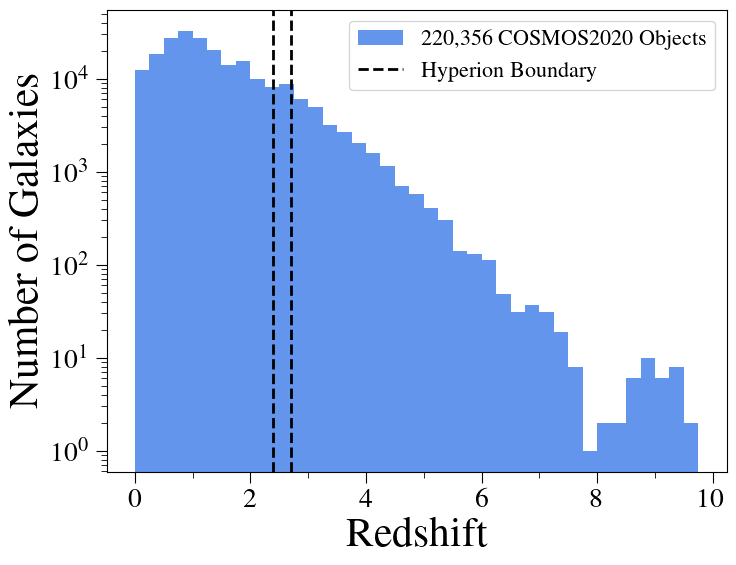}
            \caption{The redshift distribution of the 220,356 objects used in this study from the COSMOS2020 catalog after applying the cuts on object type, R.A., Dec., and IRAC magnitude described in Section \ref{sSect:C20}. The photometric redshifts were taken from \texttt{lp\_zPDF} in the COSMOS2020 catalog. We note that these are the objects that may have their redshift PDFs sampled during the Monte Carlo process (see Section \ref{sSect:zMC}). We additionally mark the approximate boundaries of Hyperion in redshift space.}
            \label{Fig:C20p_zDist}
        \end{figure}

        We imposed additional selection criteria to define our final sample of COSMOS2020 photometric sources. We limited our analysis to objects labeled as galaxies or X-ray sources in the COSMOS2020 catalog (\texttt{lp\_type} = 0 or 2), eliminating 39,653 sources. We further restricted the data to objects within the R.A. and Dec.. range of Hyperion, as determined in \citet{Cucciati+18} (see also Section \ref{Sect:GalEnvs}). After requiring $\rm 149.60^\circ \leq R.A. \leq 150.52^\circ$ and $\rm 1.74^\circ \leq Dec. \leq 2.73^\circ$, we retained 498,273 objects. Finally, to ensure completeness, we imposed a magnitude cut in the IRAC bands, requiring either $[3.6] \leq 24.8$ or $[4.5] \leq 24.8$, leaving 220,356 sources. We made no cut on redshift at this point, as a Monte Carlo routine was used later on with the reported COSMOS2020 redshift PDFs (see Section \ref{Sect:SMF}). The redshift distribution of the remaining photometric sources is plotted in Figure \ref{Fig:C20p_zDist}.

    \subsection{Ground-based spectroscopy} \label{sSect:Spec}

        \subsubsection{Wide-field surveys}\label{ssSect:FieldSpec}

            There have been a number of spectroscopic surveys which have added to the wealth of data in the COSMOS field. The zCOSMOS survey \citep{Lilly+07} was conducted using the Visible Multi Object Spectrograph (VIMOS) on the Very Large Telescope (VLT), culminating in 600 hours of observations. It consisted of two components: zCOSMOS-bright, a magnitude-limited sample spanning $0.1 < z < 1.2$, and zCOSMOS-deep, a sample of $\sim 10,000$ color-selected galaxies in the redshift range $1.4 < z < 3.0$ within the central $\rm 1\,\deg^2$ of the COSMOS field. Also taken on VIMOS was the VIMOS Ultra Deep Survey \citep[VUDS;][]{LeFevre+15}, which included spectroscopy of $\sim 10,000$ galaxies ($\sim 5000$ in the COSMOS field) with $i < 25$, spanning a redshift range of $2 < z < 6$. Additionally, the DEIMOS 10k survey \citep{Hasinger+18} using the Deep Imaging Multi-Object Spectrograph (DEIMOS) \citep{Faber+03} on the Keck II telescope took spectra of over $10,000$ objects with $i < 23$ in the redshift range of $0<z<6$, though the redshift distribution was heavily weighted to lower redshifts (see top panel of Figure \ref{Fig:Spec_Dist}). 

            To ensure the reliability of redshift measurements across these surveys, a standardized system of quality flags was used. zCOSMOS and VUDS followed a similar system for assigning quality flags to the spectroscopic redshifts, based on Section 6.5 of \citet{LeFevre+05}. The redshift quality flags from the DEIMOS 10k survey were subsequently conformed to fit the same system \citep{Lemaux+22}. Briefly, objects were assigned a flag from 0, 1, 2, 3, 4, or 9 to indicate the confidence of the redshift measurement. A complete discussion of the flag assignment and redshift reliability assessment is presented in Appendix A of  \citet{Lemaux+22}, but the redshift qualities can be summarized as follows:

            \begin{description}[style=nextline, leftmargin=10pt]
              \item[\textbf{Flags 3/4:} High confidence ($\sim 99.3\%$ reliable).] 
                     \vspace{1.5pt}
              \item[\textbf{Flags 2/9:} Moderate confidence ($\sim 70\%$ reliable).]
                     \vspace{1.5pt}
              \item[\textbf{Flag 1:} Tentative (not used in this analysis).]         
            \end{description}

            Each quality flag can be prepended with additional flags in order to indicate if the object is peculiar in some way. For example, a quality flag of 13 indicates a broadline AGN (per the prepended 1) with a secure redshift. As we are not interested in the specifics of these objects in this respect for this analysis, we ignored the prepended flags and only considered spectra which were assigned quality flags ending with 2, 3, 4, or 9 regardless of what is prepended.

            In addition to the quality flag criteria, we also required the sources have COSMOS2020 photometry so that physical properties could be derived via SED fitting (Section \ref{sSect:LP_Fits}), and are identified as either galaxies or X-ray sources in COSMOS2020. Furthermore, we only considered objects falling around Hyperion with $\rm 149.60^\circ \leq R.A. \leq 150.52^\circ$ and $\rm 1.74^\circ \leq Dec. \leq 2.73^\circ$, and which have either $[3.6] \leq 24.8$ or $[4.5] \leq 24.8$ (from COSMOS2020). After applying these cuts, we were left with 11,368 objects from the zCOSMOS survey, 4,410 objects from the DEIMOS 10k survey, and 2,129 objects from the VUDS survey. The redshift distribution of these objects is shown in the top panel of Figure \ref{Fig:Spec_Dist}.

        \subsubsection{Targeted surveys}\label{ssSect:TargetedSpec}
        
            We also drew on spectra from the C3VO program -- the Charting Cluster Construction with VUDS \citep{LeFevre+15} and ORELSE \citep{Lubin+09} survey -- which used the DEIMOS \citep{Faber+03} and Multi-Object Spectrometer For Infra-Red Exploration \citep[MOSFIRE;][]{McLean+12} instruments on Keck to follow up protocluster candidates selected with the Voronoi Monte Carlo (VMC) method \citep{Lemaux+22}. Across the fields of the Canada-France-Hawai'i Telescope Legacy Survey D1 (CFHTLS-D1), the Extended Chandra Deep Field South (ECDFS), and COSMOS, C3VO has obtained $\sim\!2000$ high-quality redshifts, yielding detailed maps of structures such as PCl J1000+0200 \citep[$z\sim 2.90$;][]{Cucciati+14}, PCl J0227-0421 \citep[$z\sim 3.31$;][]{Lemaux+14, Shen+21}, Elent\'ari \citep[$z\sim 3.33$;][]{McConachie+22, Forrest+23}, Smruti \citep[$z\sim 3.47$;][]{Forrest+17, Shah+24}, Taralay \citep[$z\sim 4.57$;][]{Lemaux+18, Staab+24}, and -- central to this work -- Hyperion.

            Details of these observations are outlined in \citet{Lemaux+22}. Briefly, the surveys covered a large on-sky area, targeting star-forming galaxies to $i < 25.3$ with Keck/DEIMOS and $H<24.0$ with Keck/MOSFIRE, and Ly$\alpha$/[O II]/H$\beta$/[O III] emitters to fainter magnitudes. Quality flags were assigned similarly to the VUDS and zCOSMOS surveys described in the previous section, though only flags 1, 3, and 4 were given. For this study, the C3VO-DEIMOS and C3VO-MOSFIRE programs contributed 280 and 159 objects, respectively. All such objects have associated COSMOS2020 photometry, are identified as either galaxies or X-ray sources in COSMOS, lie in the ranges $\rm 149.60^o \leq R.A.\leq 150.52^o$ and $\rm 1.74^o \leq Dec. \leq 2.73^o$, have either $[3.6] \leq 24.8$ or $[4.5] \leq 24.8$ (per COSMOS2020), and have spectroscopic reliability $\gtrsim 99.3\%$.

            Additionally, we included data from the Massive Ancient Galaxies at $z>3$ Near-infrared Survey \citep[MAGAZ3NE;][]{Forrest+24b}, a spectroscopic follow-up program that targeted ultra-massive galaxies ($\log_{10}(M_*/M_\odot) \gtrsim 11$ at $z>3$) using Keck/MOSFIRE in multiple fields, including COSMOS. While the galaxies from this survey do not lie within Hyperion, their spectroscopic redshifts provided valuable constraints on photometric redshift outliers. A total of 37 galaxies from MAGAZ3NE met our selection criteria.

\begin{figure}
                \centering
                \includegraphics[width=\linewidth]{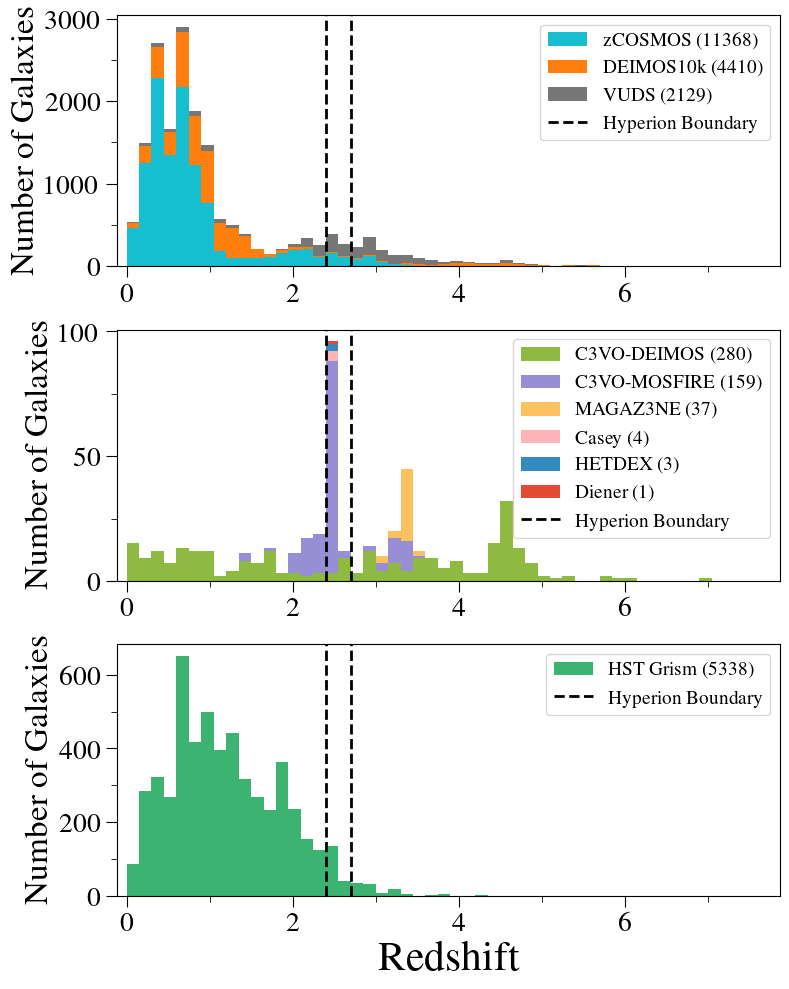}
                \caption{The distribution of spectroscopic redshifts used in this study from (top) wide-field spectroscopic surveys (section \ref{ssSect:FieldSpec}), (middle) targeted spectroscopic surveys (section \ref{ssSect:TargetedSpec}), and (bottom) the \emph{HST}-Hyperion survey (section \ref{ssSect:HST}). For each survey, we show only the galaxies that meet the imposed criteria on R.A., Dec., and IRAC magnitude described in their respective sections. Ground-based spectroscopic redshifts   all have reliabilities of $\gtrsim 70\%$, whereas \emph{HST}-grism redshifts all have reliabilities $\gtrsim 56.7\%$. The number of spectra used from each survey is reported in the legend next to the survey name. In each plot, we additionally mark the approximate boundaries of Hyperion in redshift space.
                }
                \label{Fig:Spec_Dist}
            \end{figure}

            Finally, we incorporated spectra from previous studies that targeted individual peaks of Hyperion. Specifically, we included four spectra from \citet{Casey+15}, three from \citet{Chiang+15}, and one from \citet{Diener+15}, all of which met the criteria outlined above. Further details regarding observations and data reduction can be found in the respective publications. The redshift distribution of galaxies from the various targeted spectroscopic surveys is shown in the middle panel of Figure \ref{Fig:Spec_Dist}.

    \subsection{\emph{HST} slitless spectroscopy}\label{ssSect:HST}

        In addition to the COSMOS2020 photometry and ground-based spectroscopy, we incorporated \emph{HST} slitless (grism) spectroscopy from the \emph{HST}-Hyperion Survey to greatly increase the number of secure redshifts in our sample. While details regarding target selection, data reduction, and redshift classification are provided in \citet{Forrest+25}, we briefly outline the relevant information here. \emph{HST}-Hyperion was an \emph{HST} Cycle 29 program (PI: Lemaux) which obtained 50 orbits of WFC3/G141 slitless spectroscopy with WFC3/F160W imaging in 25 pointings covering the densest peaks in Hyperion. Additionally, the survey included a reanalysis of all 3D-\emph{HST} data in the field as well \citep{Brammer+12, Momcheva+16}. In total, the observations resulted in $\sim 12,800$ spectra for objects with $m_{\rm NIR}<25$.

        Given the nature of slitless spectroscopy, there is no preselection of targets within a given field, resulting in spectra of many objects outside the redshift range of Hyperion. To vet the redshifts of each target, at least two team members independently reviewed (i) a postage-stamp image, (ii) the COSMOS2020 SED fit, and (iii) the stacked \emph{HST} spectrum. Each reviewer assigned separate numerical grades to the spectrum’s quality, the redshift fit, and the SED fit. After each classifier assessed  $\sim 2,500$ spectra, discrepancies in classification -- particularly cases where one reviewer deemed a redshift reliable while the other did not -- were resolved through discussion. In cases where a consensus was not reached, the lowest quality flag was assigned to the target.

        Following this process, a final quality flag was assigned to each object, ranging from 0 (lowest quality) to 5 (highest quality). To calibrate the reliability of the quality flags, we compared the \emph{HST}-grism redshift ($z_g$) with the ground-based spectroscopic redshift ($z_s$) for objects with both measurements. We only considered grism spectra with flags $\geq 3$ and ground-based spectra with quality flags of 3 or 4 (see Section \ref{ssSect:FieldSpec}), treating the highly reliable ground-based redshifts as the true values. We then examined the distribution of $\Delta z / (1+z_s) \equiv (z_g - z_s) / (1+z_s)$ and calculated the normalized median absolute deviation \citep[$\sigma_{\rm NMAD}$;][]{Hoaglin+83}, a measure of the distribution's width that is more robust to outliers. Denoting the median of a value $x$ as $\text{Med}(x)$, $\sigma_{\rm NMAD}$ is given as
        
            \begin{equation}\label{Eqn:NMAD}
                \sigma_{\rm NMAD} = 1.48\times \text{Med}\left(\frac{|\Delta z - \text{Med}(\Delta z)|}{1+z_s}\right).
            \end{equation}

        For the combined samples of grism redshifts with quality flags $\geq 3$, we find $\sigma_{\rm NMAD} \sim 0.0016$, corresponding to a median wavelength offset of $\sim 22\,\text{\AA}$ at $z=2.5$. Using a reliability threshold of $\Delta z / (1+z_s) < 0.004$, we determined the fraction of spectra within this cutoff for each quality flag. The reliability for flags 3, 4, and 5 is 56.7\%, 77.3\%, and 90.4\%, respectively. In total, we have grism redshifts for 5,338 objects which have reliability $\geq 56.7\%$, have associated COSMOS2020 photometry, are identified as either galaxies or X-ray sources in COSMOS, lie in the ranges $\rm 149.60^o \leq R.A.\leq 150.52^o$ and $\rm 1.74^o \leq Dec. \leq 2.73^o$, and have either $[3.6] \leq 24.8$ or $[4.5] \leq 24.8$ (per COSMOS2020). The redshift distribution of these sources is shown in the bottom panel of Figure \ref{Fig:Spec_Dist}.

    \subsection{An updated view of Hyperion}\label{sSect:HypView}

\begin{figure*}
                \centering
                \includegraphics[width=\linewidth]{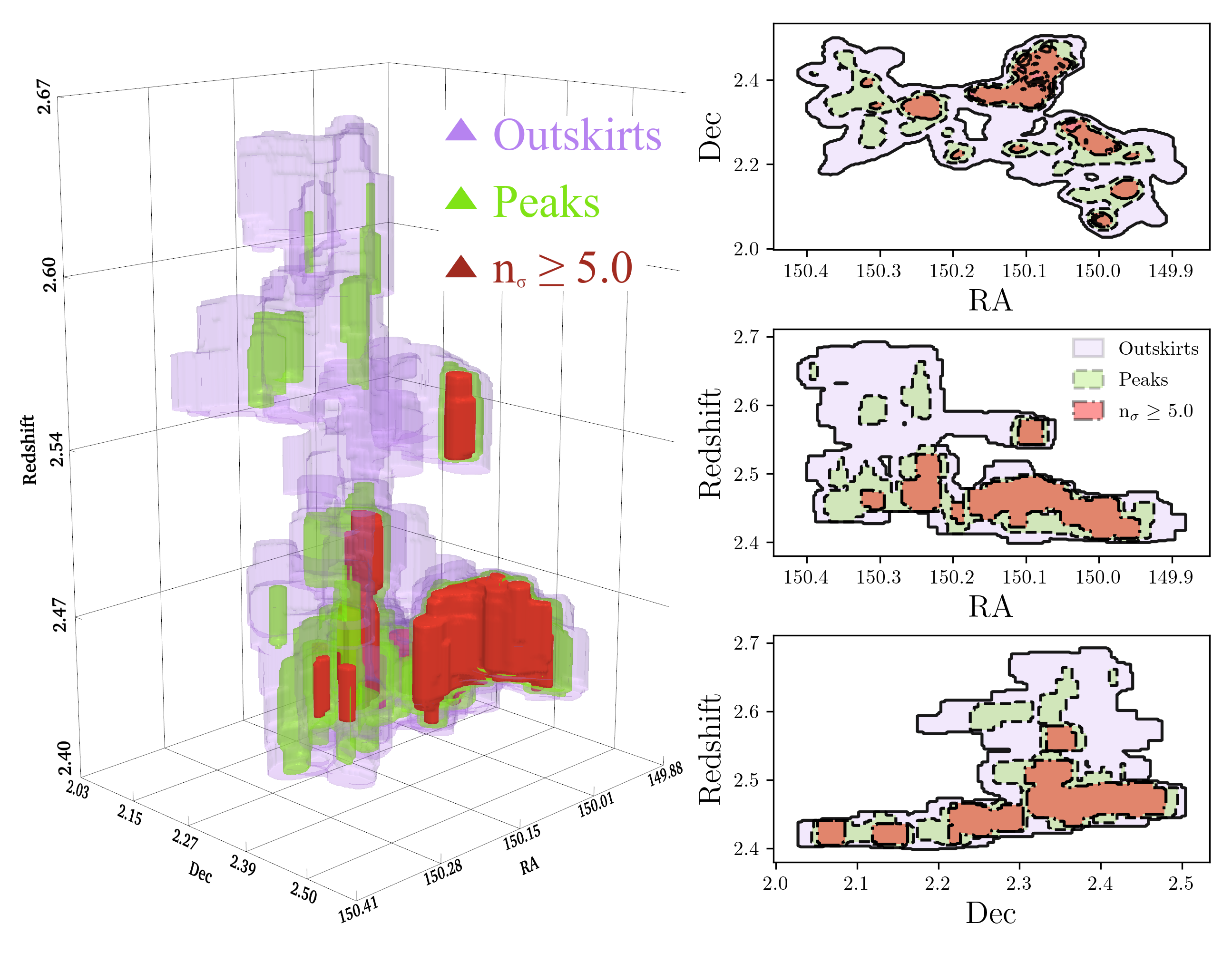}
                \caption{ Left: 3D rendering of Hyperion, as defined by the VMC technique. Right: 2D projections of Hyperion in the (top) R.A.-Dec., (middle) R.A.-$z$, and (bottom) Dec.-$z$ planes. In all panels, the contours delineate regions of differing overdensity within the VMC map. The purple and green shaded regions correspond to Hyperion's outskirts ($\rm 2.5 \leq n_\sigma \leq 4.0$) and peaks ($\rm n_\sigma \geq 4.0$), respectively, where $\rm n_\sigma$ is defined in Equation \ref{Eqn:OD}. Also shown in red are regions with $\rm n_\sigma \geq 5.0$, which generally trace the most massive peaks of Hyperion (detailed in Table \ref{Tab:Peaks}). 
                }
                \label{Fig:Hyperion}
            \end{figure*}

        Hyperion is a massive, overdense structure centered at $z\sim 2.5$, exhibiting multiple density peaks linked by filamentary bridges, as seen in Figure \ref{Fig:Hyperion}. Because such protostructures are not yet virialized, their overdensity field is highly inhomogeneous, unlike the smoother, monotonic profiles of relaxed $z\!\lesssim\!0.5$ clusters. Throughout this paper we therefore describe Hyperion in overdensity space: the highest‐density peaks (green in Figure \ref{Fig:Hyperion}) are wrapped in moderately overdense outskirts (purple in Figure \ref{Fig:Hyperion}), beyond which lies the lower-density field. Formal overdensity thresholds for the peaks, outskirts, and field are given in Section \ref{sSect:IDing_Protos}.

        Many of the peaks of Hyperion were identified as individual protostructures using a variety of techniques including submillimeter observations of tightly packed starbursting galaxies \citep{Casey+15}, the application of a friend-of-friends algorithm on deep spectroscopic surveys \citep{Diener+15}, identifying an overdensity of Ly$\alpha$ emitters \citep{Chiang+15}, and observations of extended X-ray emission around CO-emitting galaxies \citep{Wang+16}. Through the construction of a 3D Ly$\alpha$ tomographic map, \citet{Lee+16} suggested that some of the peaks may be connected in a structure they deemed Colossus. However it was not until \citet{Cucciati+18} that the grand scale of Hyperion was revealed using an overdensity map in the COSMOS field constructed from COSMOS2015 photometry \citep{Laigle+16} and spectroscopy from zCOSMOS \citep{Lilly+07} and the VIMOS Ultra Deep Survey \citep[VUDS;][]{LeFevre+15}. Based on this map, constrained cosmological simulations presented in \citet{Ata+22} suggested Hyperion will form a massive filamentary group of clusters with a final spatial extent and mass similar to that of the Coma/$\rm A\,1367$ filament \citep[][ see also results from \citet{Cucciati+23}]{Fontanelli84}.

        In this study, we used an updated overdensity map as the reference for the underlying structure of Hyperion \citep[see Section \ref{Sect:GalEnvs} for a brief summary, or a complete description in ][]{Lemaux+22}. As mapped with the VMC method, Hyperion lies in the ranges $\rm 149.88^o \leq R.A. \leq 150.41^o$, $\rm 2.03^o \leq Dec. \leq 2.50^o$, and $2.41\leq z \leq 2.68$ (shown in Figure \ref{Fig:Hyperion}). The total mass is estimated to be $\log_{10}(M/M_\odot)\sim 15.40$ and the total volume is estimated to be $\rm \sim 4.3\times10^4\,\,cMpc$. While the most overdense region of Hyperion lies at $z \lesssim 2.52$, a high-redshift offshoot appears to extend up until $z \sim 2.68$. A more detailed discussion of the structure of Hyperion, its constituent peaks, and the inclusion of the high-redshift offshoot is provided in Appendix \ref{App:HyperionStructure}.

        Leveraging the updated photometric catalog and improved spectroscopic completeness provided by the new \emph{HST} data described above, we were able to populate Hyperion with more member galaxies. The approximate locations of these members are shown in Figure \ref{Fig:HyperionMems}, although much of the following analysis relies on the full redshift probability distribution functions for each galaxy (see Section \ref{sSect:zMC}) rather than the discrete redshift estimates plotted in Figure \ref{Fig:HyperionMems}. While the relative fractions of star-forming and quiescent galaxies in Hyperion are beyond the scope of this work and will be explored in a forthcoming analysis, it is well established that the vast majority of galaxies at $z\sim 2.5$ are star-forming \citep{Weaver+23}. However, quenched galaxies -- though expected to be rare at this epoch -- were not systematically excluded from our analysis. Because quenched systems at these redshifts are typically massive \citep[$\log_{10}(M_*/M_\odot)\gtrsim 10.2$][]{Weaver+23}, their redshifts can be constrained via continuum emission in the \emph{HST}/grism data even in the absence of strong emission lines \citep{Forrest+25}. We therefore expect that, if a small population of quiescent galaxies is present in Hyperion, it is certainly represented in our sample in some capacity.

\section{Characterizing galaxy environments}\label{Sect:GalEnvs}

    Developing a quantitative method to describe the environment in which a galaxy resides is central to understanding the environmental effects on galaxy evolution. While there are a variety of methods used to quantify local environments, we employed a VMC mapping technique which has proven successful in many previous works \citep[e.g.,][]{Lemaux+17, Cucciati+18, Hung+20, Shen+21, Lemaux+22, Forrest+23, Staab+24}. This method produces a 3D grid of voxels, each with an associated overdensity value, to which galaxies are assigned. From these maps, we identified extended, connected regions of voxels exceeding a given overdensity threshold, which were taken to be protostructures. In this section, we briefly explain the VMC mapping technique before discussing the specific details of the map used in this study.

    We note that the selection criteria imposed on the galaxy populations used in both the VMC mapping and the SMF analysis are largely consistent. In particular, no explicit color cut is applied, and the only magnitude selection is based on the \emph{IRAC} magnitudes of the objects ($[3.6]\leq 24.8$ or $[4.5]\leq 24.8$), which correspond to rest-frame near-infrared emission at $z\sim 2.5$ and therefore provide a tracer of stellar mass for Hyperion members. As discussed in Section \ref{sSect:HypView}, the galaxy population at these redshifts is expected to be dominated by star-forming systems \citep{Weaver+23}. Consequently, although no explicit selection on star formation activity is imposed, the overdensity map generated via the VMC method primarily traces the spatial distribution of star-forming galaxies due to their substantially higher number density.

    \subsection{Generating overdensity maps}\label{sSect:ODMaps}
    
        \subsubsection{The Voronoi Monte Carlo method}\label{ssSect:VMC}
            The VMC algorithm leverages a combination of well-constrained spectroscopic redshifts and less-constrained photometric redshifts to statistically map the 3D spatial distribution of galaxies. Broadly speaking, a Voronoi tessellation map divides a plane into polygons, where each polygon is centered on a specific point in a given distribution. These polygons encompass all locations in the plane that are closer to their corresponding point than to any other point in the distribution, effectively partitioning the plane based on proximity. In the context of mapping galaxy density fields, the polygons partition two dimensions (R.A. and Dec.) based on proximity to the closest galaxy \citep[see Figure 3 of][]{Tomczak+17}. 

            Given the large peculiar velocities of galaxies within protostructures, the size of the uncertainties for the photometric redshifts, and the sparseness of the spectroscopic sample, this method cannot be directly applied in the redshift dimension. Instead, the tessellation maps are constructed in the 2D R.A.-Dec. plane in overlapping redshift slices. A set of Monte Carlo realizations is generated for each redshift slice of the 3D space in order to minimize the effects of photometric redshift uncertainties by better sampling the redshift PDF. In a given realization, each galaxy in the dataset is assigned a redshift based on the following logic:
            
            \paragraph{Galaxies without spectroscopic redshift:} A new redshift is drawn from an asymmetric Gaussian distribution centered on the reported photometric redshift and with lower (upper) standard deviations equal to the lower (upper) uncertainties ascribed to their redshift probability distributions.
            
            \paragraph{Galaxies with spectroscopic redshift:} Each quality flag has an associated likelihood and uncertainty (e.g., $0.70_{-0.06}^{+0.04}$ for flags X2/X9), and threshold for the realization is drawn from an asymmetric Gaussian with the associated mean and dispersion. A random number is also drawn from a uniform distribution ranging from 0 to 100. If the number is less than or equal to the threshold associated with the quality-flag of the spectroscopic redshift (e.g., $\lesssim 70_{-6}^{+4}$ for flags X2/X9), the reported spectroscopic redshift is kept. Otherwise, a new redshift is drawn from the photometric redshift distribution, as described above.\\
            
            Only the galaxies assigned redshifts in the given redshift slice are kept for each realization, and a 2D Voronoi tessellation is performed to create a 2D polygon map. The density associated with a polygon, $\Sigma_{\alpha,\delta,z}$, is then inversely proportional to its area in $\rm Mpc^2$. This is remapped onto a consistent grid of voxels in R.A.-Dec. space separated by 75 pkpc, where each voxel has the density of the polygon enclosing its center in the given Monte Carlo realization.

        \subsubsection{Applying the VMC method}\label{ssSect:MeasureOD}

            The VMC maps used in the present study are described in \citet{Lemaux+22}, and thus we give only a brief description here. The maps were generated using the COSMOS2015 photometry \citep{Laigle+16} and associated field spectroscopic surveys (described in Section \ref{ssSect:FieldSpec}). Targeted spectroscopic surveys were used primarily to improve statistics at redshifts higher than Hyperion in an effort to mitigate biases in overdensities of targeted, intermediate-redshift protostructures. The VMC process, as described above, was performed over the entire COSMOS footprint for redshifts $2\leq z \leq 5$. The redshift bins had widths of $7.5 \,\,\text{Mpc}$, with overlaps of 90\% between adjacent redshift slices. A total of 100 MC realizations were performed for each redshift slice to generate the map, using only those galaxies that fell within the redshift bin and with magnitudes $[3.6] < 24.8$ for the density calculations in a given realization. Afterwards, each voxel was assigned a density $\Sigma_{\alpha,\delta,z}$ equal to the median density of that voxel across all MC realizations.

            The galaxy density measurement found via VMC mapping is subject to differences in the quality of data used, magnitude cuts, and other observational constraints, as well as the true average density of the Universe varying with redshift \citep{Hung+20}. Because we are interested in quantifying galaxy evolution with respect to the surrounding environment and comparing this to the evolution at other redshifts, it is more useful to use the local overdensity, $\log_{10}(1+\delta_{\rm gal})$ -- rather than just density -- as an environmental metric. Given the density of a voxel $\Sigma_{\alpha,\delta,z}$, and the average density for the redshift slice containing the voxel, $\langle\Sigma_z\rangle$, the overdensity of the voxel, is defined as
            
            \begin{equation}\label{Eqn:OD}
                \log_{10}(1+\delta_{\rm gal}) \equiv \log_{10}\left(1+\frac{\Sigma_{\alpha,\delta,z}}{\langle\Sigma_z\rangle}\right).
            \end{equation}

\begin{figure*}
                \centering
                \includegraphics[width=\linewidth]{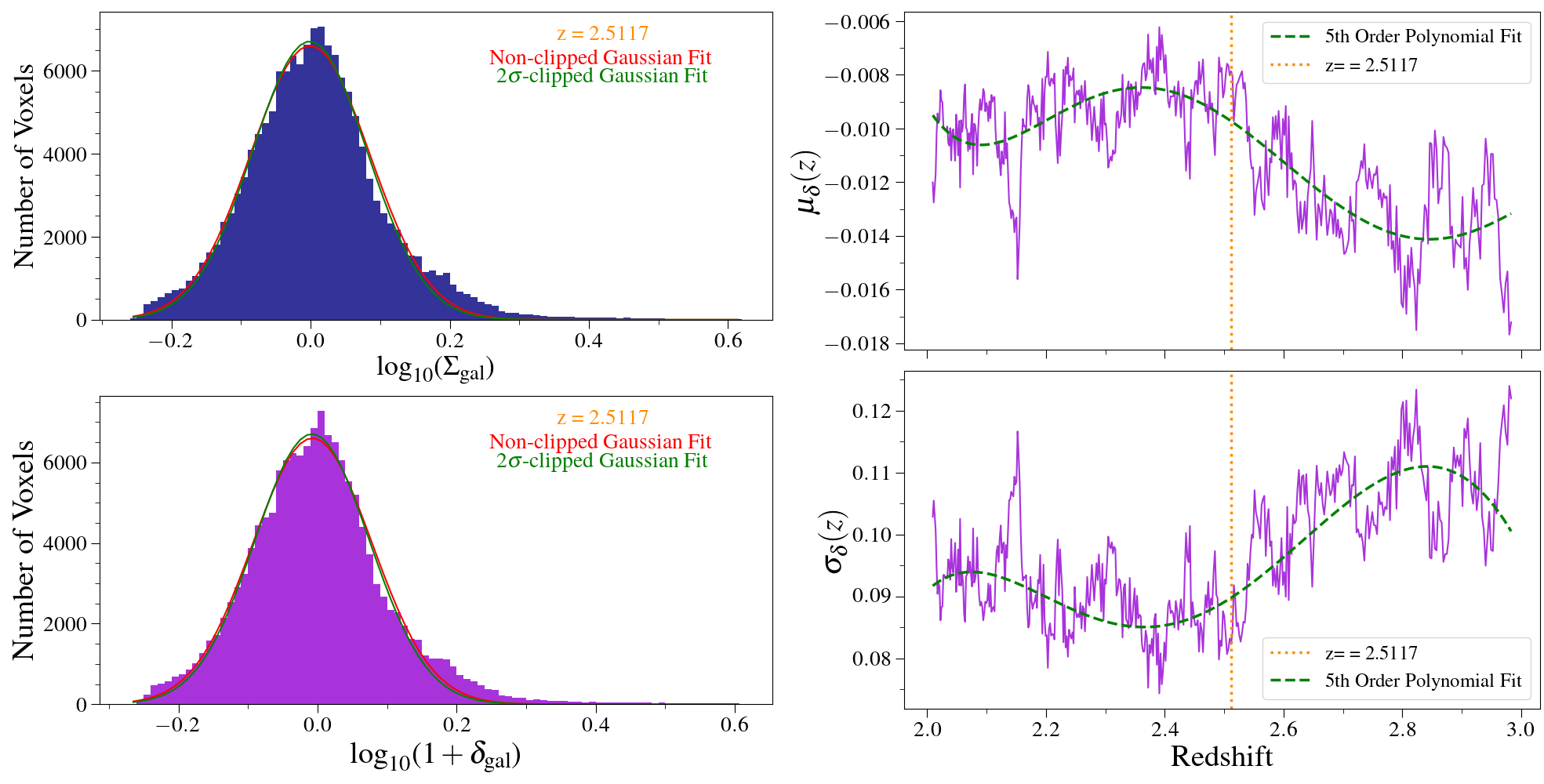}
                \caption{Left: Distributions (with median set to zero) of log-density (top), and local overdensity (bottom) for all of the voxels in the redshift bin centered on $z\sim 2.51$. The Gaussian and $2\sigma$ clipped Gaussian (red and green, respectively) are fit over both distributions. Right: Distribution of the mean overdensity (top) and standard deviation of overdensity (bottom) as a function of redshift. The functions are fit with a fifth-order polynomial (green line) in order to smooth over small fluctuations in redshift space.}
                \label{Fig:OD_Dist}
            \end{figure*}

            In order to find $\langle\Sigma_z\rangle$, the outer 10\% of voxels in each redshift slice were first trimmed to mitigate edge effects. The remaining 2D density distribution in the redshift slice, $\Sigma_z$, was then smoothed with a 2D-Gaussian kernel with standard deviations of three pixels in the two transverse dimensions (R.A. and Dec.). Additionally, a boxcar smooth was performed in the redshift dimension with a width of three slices. Since $\Sigma_z$ roughly follows a log-normal distribution, a Gaussian was fit to the distribution of $\log_{10}(\Sigma_z)$ in each redshift slice. Finally, a $2\sigma$-clipping was applied and another Gaussian was fit, resulting in a mean $\mu_\Sigma(z)$ and standard deviation $\sigma_\Sigma(z)$ for each redshift slice (top left panel of Figure \ref{Fig:OD_Dist}). The mean density in the redshift-bin is related to the mean and standard deviation of the log-normal distribution via $\langle\Sigma_z\rangle = 10^{\mu}\cdot\exp(2.652\sigma^2)$.

            Once the average density is found in each redshift slice, the overdensity can be calculated for each voxel in the slice from Equation \ref{Eqn:OD}. As above, the distribution of \OD in the slice was fit with a $2\sigma$-clipped Gaussian resulting in a $\mu_\delta(z)$ and $\sigma_\delta(z)$ for that slice (bottom left panel of Figure \ref{Fig:OD_Dist}). It should be noted that, given that protostructures are extended in redshift space over multiple redshift slices, it is possible that an entire redshift slice is over- or underdense. Thus, a fifth-order polynomial was fit to $\mu_{\delta,z}$ and $\sigma_{\delta,z}$ to better characterize the general variation across redshift (see right panels of Figure \ref{Fig:OD_Dist}), resulting in measurements for the parameters as a function of redshift, $\mu_{\delta,5}(z)$ and $\sigma_{\delta,5}(z)$. 

            With these in hand, we now have a metric for characterizing the local overdensity in each voxel of the VMC map by representing it as
            \begin{equation} \label{Eqn:PolyFit}
                \log_{10}(1+\delta_{\rm gal}) = \mu_{\delta,5}(z) + n_\sigma \sigma_{\delta,5}(z).
            \end{equation} 
            With this form, the local overdensity is parameterized entirely by $n_\sigma$, or the number of $\sigma_{\delta,5}$ above $\mu_{\delta,5}$ the local overdensity for a given voxel. This metric is robust over cosmic fluctuations given the polynomial fit, as well as observational constraints given the MC process used to generate the underlying \OD distribution.

    \subsection{Identifying protostructures}\label{sSect:IDing_Protos}

        Due to their inherently inhomogeneous overdensity profiles, the precise spatial extent of protostructures can be difficult to define and somewhat ambiguous. Therefore, we adopted criteria intended to provide an operational definition within the VMC framework which is both physically motivated and consistent with previous works. Specifically, we defined a region to be a ``protostructure" if it satisfied all three of the following criteria:
            \begin{enumerate}[label=(\roman*),leftmargin=*]
                \item it was a contiguous set of voxels with overdensity $n_\sigma \geq 2.5$;
                \item it contained at least one voxel with $n_\sigma \geq 4.0$, corresponding to an overdensity peak;
                \item the voxels together enclosed a total mass $M_{\rm tot} \geq 10^{13}\,M_\odot$.
            \end{enumerate}

        Criterion (i) sets the outer boundary of Hyperion at an overdensity threshold comparable to that used in \citet{Cucciati+18}. Although \citet{Cucciati+18} adopted a slightly lower threshold of $n_\sigma > 2.0$, both choices correspond to $\log_{10}(1+\delta_{\rm gal}) \sim 0.21$ at $z\sim 2.5$ due to the use of different VMC maps. As a result, the structure identified here as Hyperion has a similar projected spatial extent -- though a somewhat different redshift distribution -- to that mapped in \citet{Cucciati+18}. Since this definition also motivated the confined cosmological simulations of Hyperion presented in \citet{Ata+22}, their results likewise imply the structure we identify as Hyperion will evolve into a massive filamentary supercluster comparable in scale and mass to the Coma/$\rm A\,1367$ system \citep{Fontanelli84}.

        The $n_\sigma \geq 4.0$ threshold appearing in criterion (ii) sets the boundary between overdensity peaks and lower-density outskirts within Hyperion and was chosen to facilitate comparison with other studies, particularly the analysis of the Elent\'ari proto-supercluster presented in \citet{Forrest+24}. Although Elent\'ari also resides in the COSMOS field, it lies at $z\sim3.3$, making a direct comparison in terms of $\delta_{\rm gal}$ unphysical. We therefore adopted an identical $n_\sigma$-threshold to define peaks, as this metric is more readily comparable across redshift. In our maps, $n_\sigma \geq 4.0$ corresponds to $\log_{10}(1+\delta_{\rm gal}) \sim 0.35$ at $z\sim2.5$.
        
        Finally, criterion (iii) roughly corresponds to a condition imposed on the average overdensity of the protostructure and the number of voxels from which it is composed. The mass of the voxels is estimated by using \citep[from][]{Steidel+98}
        \begin{equation}\label{Eqn:PeakMass}
            M_{\rm tot} = \bar{\rho} \,V\left(1 + \frac{\langle\delta_{\rm gal}\rangle}{b} \right),
        \end{equation}
        where $\bar{\rho}$ is the mean comoving density of the universe, $\langle\delta_{\rm gal}\rangle$ is the mean overdensity of the voxels defining the peak, $V$ is the combined comoving volume of the voxels, and $b$ is the bias factor, \citep[2.55, as derived at $z\sim 2.5$ by][]{Durkalec+15}. It should be noted that, while this mass estimate may be rough, in this work it is used only as a method for differentiating between regions that are overdense because they are potentially protostructures, and regions that are overdense simply due to cosmological perturbations in density. Given the robustness of our field sample (see Figure \ref{fig:TotalSMF}) and the functional form of Equation \ref{Eqn:PeakMass}, any uncertainties in the bias factor will have no meaningful effects on the conclusion of this paper.

        With this framework established, Hyperion was defined as the most massive structure in the COSMOS field. By considering only voxels within Hyperion above higher $n_\sigma$ thresholds, we could further isolate its internal overdensity structure. Using the $n_\sigma \geq 4.0$ threshold appearing in criterion (ii), we defined the ''peaks" of Hyperion to be contiguous sets of voxels within Hyperion with $n_\sigma \geq 4.0$. The ``outskirts" of Hyperion were then defined as the remaining voxels associated with Hyperion that lie in the overdensity range $2.5 \leq n_\sigma < 4.0$. A more detailed description of the structure of Hyperion is provided in Appendix \ref{App:HyperionStructure}.

        Finally, we note that the results of this study are not sensitive to the precise choice of overdensity thresholds used to define protostructures in the VMC maps. Following the framework presented in Section \ref{Sect:SMF}, we have verified that the SMF of Hyperion varies smoothly as a function of the overdensity threshold used to define it (i.e., as a function of $n_\sigma$ in criterion (i)), up to $n_\sigma \gtrsim 4.5$, beyond which the available sample size of galaxies becomes too small for robust statistical comparison.

    \subsection{Creating a field sample}\label{sSect:IDing-Fields}

        To gauge the effect of overdense environments, we need a control set of galaxies that have grown in lower-density regions. We therefore drew from two field slices that bracket Hyperion, $2.15 \le z \le 2.25$ and $2.80 \le z \le 2.90$. These ranges kept look-back-time differences small while minimizing the chance that Hyperion members scattered into the field sample (or vice-versa).
        
        Defining the field solely as voxels with $n_\sigma < 2.5$ would require excluding every voxel with $n_\sigma \ge 2.5$, erasing genuine large-scale fluctuations caused by cosmic variance and yielding an unrealistically uniform field. Instead, we removed only those galaxies that belong to a VMC-identified protostructure at redshifts $2 \leq z \leq 3$, adopting the definition of a protostructure defined in the previous section.

\section{Building the stellar mass functions}\label{Sect:SMF}

    Despite our extensive spectroscopic data within Hyperion, the total sample is still dominated by galaxies with relatively poorly constrained photo-$z$s (compared in Figures \ref{Fig:C20p_zDist} and \ref{Fig:Spec_Dist}). In theory, photometric galaxies which fall outside of the range of Hyperion could in fact be located in one of its peaks, changing the SMF of Hyperion, and vice-versa. To extract as much information as we can from our sample we must correctly quantify our uncertainty regarding which peak (if any) a galaxy is located in by correctly accounting for each galaxy's redshift uncertainty. However, as the stellar masses are derived via SED-fitting, the redshift of a galaxy is intricately tied to the estimated stellar mass of the galaxy. Thus, a redshift uncertainty results not only in an uncertainty in the 3D location (and thus overdensity via the VMC maps) of a galaxy, but also in the galaxy's stellar mass. 

    Rather than making assumptions about how the stellar mass changes as a function of redshift, we instead chose to sample the redshift distributions of our data via a Monte Carlo (MC) process, and refit the galaxy SED at each sampled redshift. In this section, we describe how we generated the MC realizations, and how each galaxy in a given realization was fit with a new stellar mass estimate generated in the process, before finally creating the SMFs based on the resulting galaxy distributions. It should be noted that this process was performed separately from the MC process used to generate the overdensity map (Section \ref{sSect:ODMaps}). In what follows, we sampled only the redshifts of the galaxies and assigned an overdensity based on the already-generated VMC map.

    \subsection{Probing redshift distributions}\label{sSect:zMC}

        In order to account for redshift uncertainties, we created a suite of 100 Monte Carlo realizations in which we attempted to accurately sample the redshift PDFs of our data. Broadly speaking, a given object in our sample has a \pz, \sz, \gz, or some combination of the three. Therefore, the goal was, for galaxies with more than just a \pz, to use the more secure redshift measurement (either \sz or \gz) a number of times which is proportional to the reliability measurement, given by the quality flag (see Section \ref{Sect:Data}).

        For a given galaxy in a given MC iteration, the logic for choosing a redshift depended on the combination of redshift measurements available for that galaxy. For a galaxy with all three types of redshift measurements, we potentially drew a new redshift from three different PDFs. Given each galaxy in our sample has COSMOS2020 photometry, they also have associated redshift PDFs from COSMOS2020 based on fits in \texttt{LePhare}. If the redshift in the given MC iteration was based on the \pz, the redshift was drawn randomly from the COSMOS2020 PDF, $p(z)$. If based on the \gz, redshifts were randomly drawn from a Gaussian distribution, $g(z)$, centered on the measured redshift from the grism spectroscopy, $z_g$, and with a width equal to $\sigma_g = 46\cdot(1+z_g)/14100$ (motivated in Section \ref{ssSect:HST}). Finally, if based on the \sz, redshifts were simply taken to be the measured spectroscopic redshift, $z_s$. Thus, we outline how the random redshift, $z_{\rm MC}$, is drawn based on whether the galaxy has a \pz ($z_p$), \sz ($z_s$), \gz ($z_g$), or some combination of the three. 

        \paragraph{Only $z_p$}: 
        The vast majority of galaxies ($\sim 90.3\%$) only have a \pz, which is reported in COSMOS2020. For these galaxies, we simply drew $z_{\rm MC} \sim p(z)$ for each MC iteration from the redshift PDF provided in COSMOS2020 from a \texttt{LePhare} fit.
        
        \paragraph{$z_p$ and one of $z_s$ or $z_g$}: 
        In this case, the galaxy has COSMOS2020 photometry and either a $z_g$ or $z_s$. The \sz or \gz also has an associated reliability flag corresponding to some confidence in the redshift, $qf$ (described in Section \ref{Sect:Data}). For each MC iteration, a random number, $\mathcal{R}$, was drawn from a uniform distribution on the half-open interval $[0,1)$. If $\mathcal{R} \geq qf$, then we took $z_{\rm MC} \sim p(z)$. Otherwise, we took $z_{\rm MC} = z_s$ or $z_{\rm MC} \sim g(z)$, depending on whether the additional redshift is a \sz or \gz.
        
        \paragraph{$z_p,\,\,z_s$ and $z_g$}: 
        Some galaxies in our data have all three types of redshift measurements. In this case, the first step was to compare the confidence levels corresponding to the quality flags of the additional redshift measurements, which we call $qf_{\rm s}$ and $qf_{\rm g}$ for the \sz and \gz, respectively. For clarity, we define the quantity $\mathcal{Q} \equiv qf_{\rm max} + qf_{\rm min}\cdot (1- qf_{\rm max})$, corresponding to a $qf_{\rm min}$-fraction of the way between $qf_{\rm max}$ and 1. That is, if $qf_{\rm max} = 0.95$ and $qf_{\rm min} = 0.7$, (corresponding to 95\% and 70\% confidences in the additional redshift measurements) then $\mathcal{Q} = 0.95 + 0.7\cdot(0.05) = 0.985$, or 70\% of the way between 0.95 and 1. This way, if the less reliable of the additional redshifts is selected in a given MC iteration, the remaining interval $[qf_{\rm max}, 1)$ is weighted according to the confidence associated with that less reliable measurement.\\
        
        With this defined, a random number, $\mathcal{R}$, was drawn from a uniform distribution on the half-open interval $[0,1)$, and a redshift was assigned depending on which of the additional redshift measurements had a higher confidence. In the case $qf_{\rm max} = qf_{\rm s} > qf_{\rm g} = qf_{\rm min}$ (the \sz is more reliable than the \gz), a redshift was assigned according to
        \begin{equation*}
            \begin{cases}
                \mathcal{R} < qf_{\rm max} & \Rightarrow  z_{\rm MC} = z_s \\
                qf_{\rm max} \leq \mathcal{R} < \mathcal{Q} &\Rightarrow z_{\rm MC} \sim g(z) \\
               \mathcal{Q} \leq \mathcal{R} < 1 &  \Rightarrow z_{\rm MC} \sim p(z)
            \end{cases}.
        \end{equation*}
        Conversely, if $qf_{\rm max} = qf_{\rm g} > qf_{\rm s} = qf_{\rm min}$ (the \gz was more reliable than the \sz), a redshift was assigned according to
        \begin{equation*}
            \begin{cases}
                \mathcal{R} < qf_{\rm max} & \Rightarrow  z_{\rm MC} \sim g(z) \\
                qf_{\rm max} \leq \mathcal{R} < \mathcal{Q} &\Rightarrow z_{\rm MC} = z_s \\
               \mathcal{Q} \leq \mathcal{R} < 1 &  \Rightarrow z_{\rm MC} \sim p(z)
            \end{cases}.
        \end{equation*}
        
        We applied this process to all $220,356$ galaxies in our sample, shown in Figure \ref{Fig:C20p_zDist}. In each MC iteration, we chose a redshift for all $220,356$ galaxies, of which $\sim 9.7\%$ additionally had a spectroscopic or grism redshift. For future steps in a given MC iteration we only considered galaxies which fell in the redshift range $2<z<3$.

    \subsection{Fitting stellar mass}\label{sSect:LP_Fits}

        For each Monte Carlo realization, we refitted the stellar mass of every galaxy based on its MC-assigned redshift. In order to refit the masses, we used \texttt{LePhare} with similar parameters to those used in the COSMOS2015 and COSMOS2020 catalogs \citep{Ilbert+15, Laigle+16, Weaver+22}. Since we are working with high-redshift data and are interested in a statistical study of a broad population of galaxies, we used only galaxy SED templates, excluding stellar and QSO templates. In this section, we briefly discuss the parameters used for the SED fits.

        The SED fitting was performed using the flux densities reported in COSMOS2020. As described in Section \ref{sSect:C20}, for a given a band \texttt{XXXX}, we used the aperture-to-total corrected flux density reported under \texttt{XXXX\_FLUX\_APER3}, and its error from the COSMOS2020 Classic catalog. When available, we used fluxes from 35 bands including all ground-based observations, \emph{Spitzer}/\emph{IRAC} channels 1 and 2, and NUV and FUV data from GALEX. We applied a systematic magnitude offset to the reported photometry (\texttt{APPLY\_SYSSHIFT} in \texttt{LePhare}) given in Table 3 of \citet{Weaver+22}. 

        To fit for the physical properties of the galaxies, we used a set of twelve simple stellar population (SSP) templates from \citet{Bruzual+03}. For all twelve templates, we assumed a \citet{Chabrier03} initial mass function, in which eight templates had exponentially declining and four had a delayed star formation histories. The former had SFR e-folding time periods in the range $\rm 0.01\,\,Gyr\leq\tau\leq 30\,\,Gyr$ and the latter had either $\rm\tau = 1\,\,Gyr$ or $\rm\tau = 3\,\,Gyr$. Models assumed either half-solar or solar metallicities. We assumed two different models for attenuation for all but the templates with the lowest SFRs: one from starbursting galaxies \citep{Calzetti+00} and one assuming a wavelength dependence of $\lambda^{0.9}$ \citep{Arnouts+13}. We assumed color-excesses of $E(B-V) = 0.,\,0.05,\,0.1,\,0.2,\,0.3,\,0.5,\, \text{or }1$. Dust emission was accounted for by adding flux contributions from templates via \citet{Bethermin+12} to the SSP templates. Emission lines were added to the templates using the \texttt{EMP\_UV} option in \texttt{LePhare}, which uses relations from \citet{Kennicutt98} to relate UV luminosity to an SFR, which in turn defines an H$\alpha$ luminosity and subsequently other line-luminosities based on pre-defined flux ratios \citep{Ilbert+09}. Emission lines were allowed to vary from the expected ratios by factors of 0.25, 0.5, 1, and 2, and the models were evaluated at 43 different galactic ages.

\begin{figure}
            \centering
            \includegraphics[width=\linewidth]{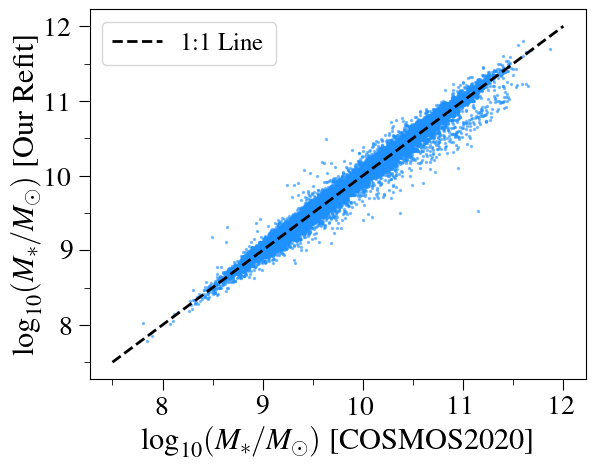}
            \caption{We show the stellar masses of galaxies in the COMSMOS catalog (\texttt{lp\_MASS\_MED}) compared to our \texttt{LePhare} outputs as a consistency check for our SED fitting parameters. The galaxies tested are those in Figure \ref{Fig:C20p_zDist} in the redshift range $2 \leq z \leq 3$ (as reported from \texttt{lp\_zPDF} in COSMOS2020). The fits are consistent with the 1:1, indicating our SED fitting parameters are broadly consistent with those used in COSMOS2020.
            }
            \label{Fig:C20_MassFit}
        \end{figure}

        To ensure consistency of our fits with those reported in COSMOS2020, we refitted a subset of relevant galaxies in the COSMOS2020 catalog and compared the resulting stellar masses. Specifically, we refitted the $33,097$ galaxies which fall in the redshift range $2 \leq z \leq 3$ in Figure \ref{Fig:C20p_zDist}, fixing the redshift to the value given by \texttt{lp\_zPDF} in COSMOS2020. In Figure \ref{Fig:C20_MassFit}, we compare the stellar masses reported in COSMOS2020 via \texttt{lp\_mass\_med} with those from our refit. The fits are broadly consistent with 1:1, indicating the parameters used for SED fits in this study replicate those used in COSMOS2020 \citep{Weaver+22} for the range of stellar masses we consider in this study.

    \subsection{Creating and fitting the stellar mass functions}\label{sSect:Build_SMFs}
    
        We constructed a stellar mass function for both Hyperion and the field in each of the 100 MC realizations. We restricted our analysis only to those galaxies which have $\log_{10}(M_*/M_\odot)\geq 9.5$ to ensure a more complete sample. For each MC realization, we assigned galaxies to either Hyperion or the field sample as defined in Sections \ref{sSect:IDing_Protos} and \ref{sSect:IDing-Fields}, respectively. We then counted the number of galaxies in each stellar mass bin, normalizing by the bin width and the volume of the corresponding population. We find no significant difference between the high- and low-redshift field samples, and therefore combined them into a single field SMF.
        
        The volumes of Hyperion and its constituent overdensity peaks were computed as the sum of the volumes of the voxels representing them (see Section \ref{ssSect:VMC}). For the field, we took the volume to be the rectangular prism that spans the full ranges in R.A., Dec., and redshift for all galaxies, excluding the volumes of any protostructures within it (see Section \ref{sSect:IDing_Protos}).

\begin{figure}
            \centering
            \includegraphics[width=1\linewidth]{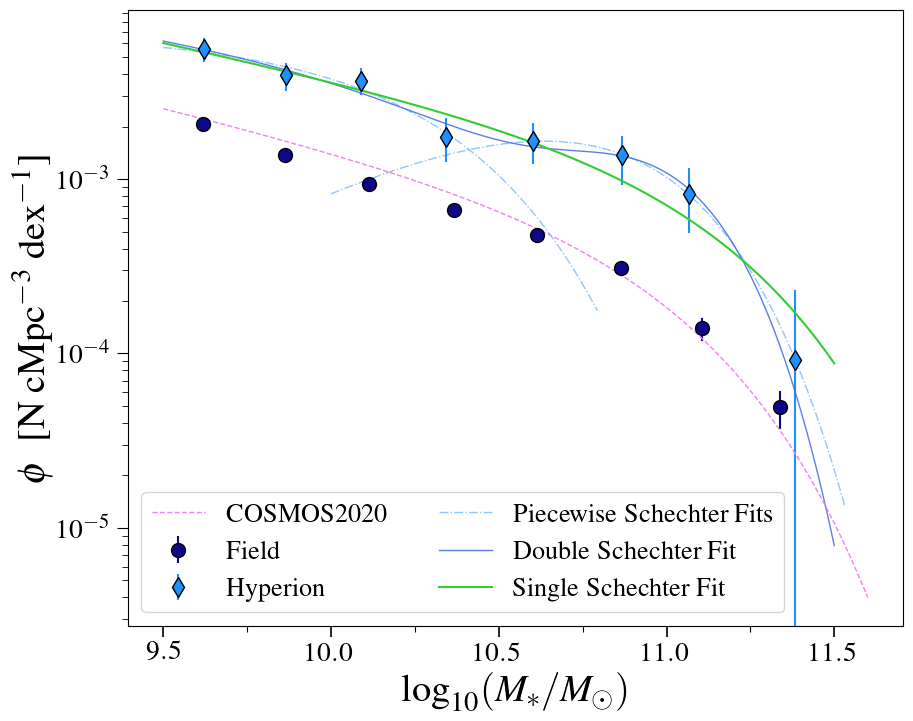}
            \caption{We show the total SMF of Hyperion as well as the coeval field. Additionally, we show the single-Schechter fit of the COSMOS2020 fit for $2.5 < z < 3.0$ \citep[magenta line;][]{Weaver+23}. Finally, we show the best-fit single-Schechter (green) and double-Schechter (blue) functions to the Hyperion SMF (see parameters in Table \ref{Tab:Sch}).}
            \label{fig:TotalSMF}
        \end{figure}

        In Figure \ref{fig:TotalSMF}, we report the median SMF across the 100 MC realizations for the Hyperion and the field sample, retaining only mass bins with at least one galaxy on average per realization. Upon averaging the MC realizations, we find Hyperion contains $207_{-9}^{+9}$ galaxies with $\log_{10}(M_*/M_\odot)\geq 9.5$, compared to an average of $3518_{-48}^{+49}$ in the field. Uncertainties are reported as asymmetric error bars, combining (in quadrature) the Poisson error across all realizations and the spread between the median and the 16th/84th percentiles.

        We additionally fitted two forms of the Schechter function \citep{Schechter76} to the Hyperion SMF. The single-Schechter form is given by
        \begin{equation}\label{Eqn:Sch}
        \begin{split}
            n_{\rm gal} &= \Phi\,\,d\log_{10} M \\
            &= \ln(10)\Phi^*\,
            e^{-10^{\log_{10} M - \log_{10} M^*}}  \\
            & \quad \quad\times (10^{\log_{10} M - \log_{10} M^*})^{\alpha+1} \,\,d\log_{10} M,
        \end{split}
        \end{equation}
        which describes a power-law slope $\alpha$ at low mass and an exponential cutoff above the characteristic mass $M^*$. However, many studies at lower redshifts \citep[e.g.,][]{Peng+10} find that a double-Schechter function provides a better fit:
        \begin{equation}\label{Eqn:DSch}
        \begin{split}
            n_{\rm gal} &= \Phi\,\,d\log_{10} M \\
            &= \ln(10)\, e^{-10^{\log_{10} M - \log_{10} M^*}}  \\
            &\quad\times \bigl[\Phi^*_1(10^{\log_{10} M - \log_{10} M^*})^{\alpha_1+1} \\
            &\qquad + \Phi^*_2(10^{\log_{10} M - \log_{10} M^*})^{\alpha_2+1} \bigr]\,\,d\log_{10} M.
        \end{split}
        \end{equation}
        Qualitatively, this function is described by a single characteristic mass, $M^*$, above (below) which the SMF is described by a normalization $\Phi^*_1$ ($\Phi^*_2$) and a power-law slope $\alpha_1$ ($\alpha_2$).

        The best-fit Schechter-functions are shown in Figure \ref{fig:TotalSMF}, along with a single-Schechter fit to the COSMOS2020 field sample in the range $2.5 < z \leq 3$, based on only photometry \citep[Table C.1 in ][]{Weaver+23}. Best-fit parameters and respective reduced $\chi^2$-values for the single- and double-Schechter fits are given in Table \ref{Tab:Sch}. We additionally report the Bayesian information criterion \citep[BIC;][]{Schwarz78}, calculated via
        \begin{equation}\label{Eqn:BIC}
            \text{BIC} = k\ln(n) + \chi^2,
        \end{equation}
        where $k$ is the number of parameters in the model and $n$ is the number of data points. The BIC is constructed to penalize models with more free parameters, and is thus useful for comparing models of different complexity, where lower BICs are typically better as high BICs may indicate a worse model or overfitting. In practice, the three parameters of the Schechter function -- $\alpha$, $\Phi^*$, and $M^*$ -- can be degenerate with one another, thus making interpretation of the parameters and their errors difficult. Covariances between these parameters and their interpretations are discussed in Appendix \ref{App:Schech}.

\begin{figure}
            \centering
            \includegraphics[width=1\linewidth]{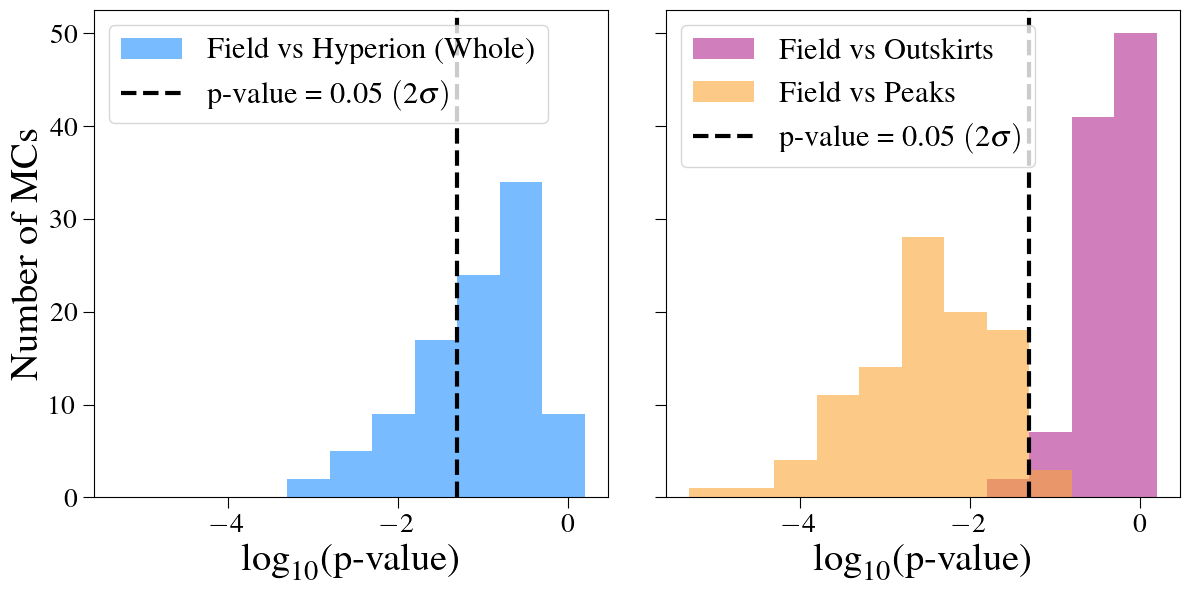}
            \caption{The results of two sample KS tests that were run for each of the 100 MC realizations in which we compare the mass distribution of the field to that of (left) Hyperion as a whole and (right) Hyperion's outskirts (pink) and overdensity peaks (orange). We find $33\%$ of MCs in which Hyperion as a whole is statistically different from that of the field ($p<0.05$), compared to $2\%$ and $97\%$ for the Hyperion outskirts and peaks, respectively. Thus, the underlying mass distribution for Hyperion's peaks are significantly different from the field for most of the MC realizations, compared to Hyperion's outskirts, which show no significant difference.}
            \label{fig:KS_Test}
        \end{figure}

        In order to compare the SMF of Hyperion to that of the field, we performed a two-sample Kolmogorov-Smirnov (KS) test -- a nonparametric method of testing whether two samples are drawn from the same underlying distribution. In this case, we treated each of the 100 MC realizations separately and performed a two-sample KS test comparing the stellar masses of the galaxies in the field to those in Hyperion. The resulting distribution of $p$-values for the 100 MC iterations is shown in the left panel of Figure \ref{fig:KS_Test}, in which the null hypothesis is that the underlying mass distributions are the same for the field and Hyperion. In total, we find 33/100 MCs with a $p$-value $< 0.05$.
        
\begin{figure*}
            \centering
            \includegraphics[width=1\linewidth]{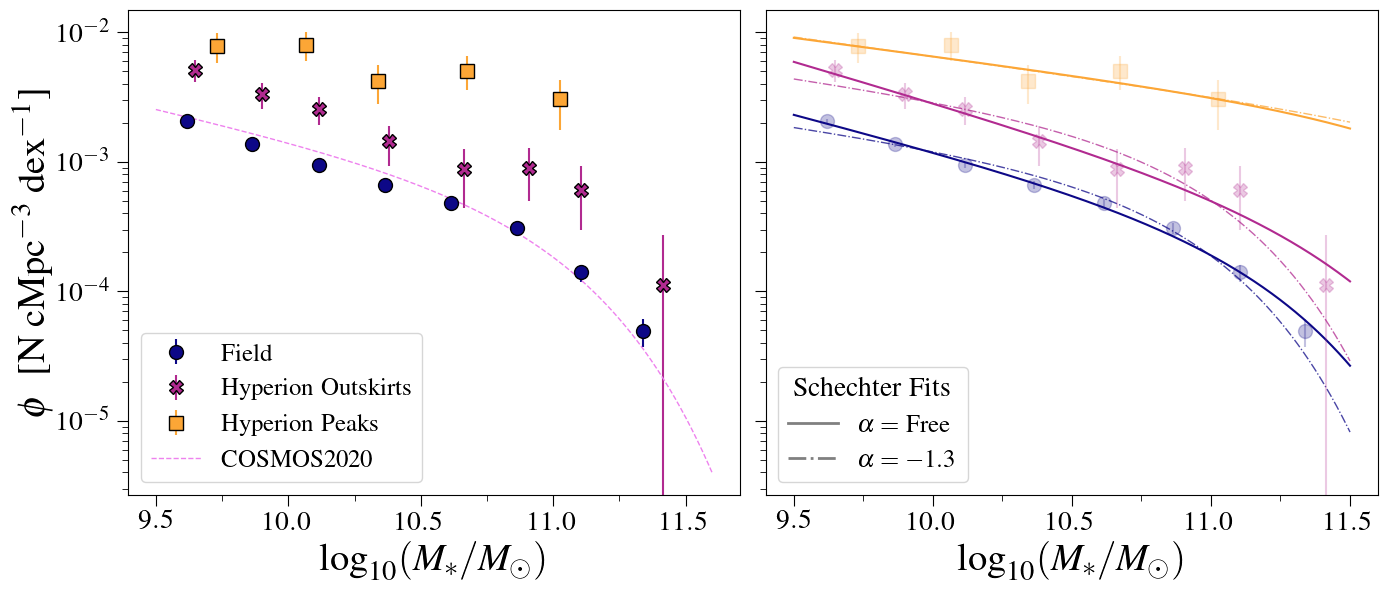}
            \caption{Left: SMFs of the outskirts ($2.5 \leq n_\sigma < 4.0$) and overdensity peaks ($4.0 \leq n_\sigma$) of Hyperion. The coeval field and COSMOS2020 fit are identical to Figure \ref{fig:TotalSMF}. Right: Best-fit Schechter functions for the outskirts and overdensity peaks of Hyperion, and the coeval field. The solid lines correspond to single-Schechter fits allowing all three parameters to vary freely; the dot-dashed lines correspond to fits fixing the power-law slope to $\alpha = -1.3$. The best-fit parameters for each fit are given in Table \ref{Tab:Sch}.}
            \label{fig:ODSMF}
        \end{figure*}

        In addition to the SMF for Hyperion as a whole, we also built SMFs for different discrete overdensity thresholds within Hyperion. Specifically, we used the MC iterations to build SMFs for galaxies within the outskirts of Hyperion ($2.5 \leq n_\sigma < 4.0$) and the peaks of Hyperion ($4.0 \leq n_\sigma$), where $n_\sigma$ is given by Equation \ref{Eqn:PolyFit}. The resulting SMFs are shown in the left panel of Figure \ref{fig:ODSMF}. For the field sample and each overdensity bin within Hyperion, we fitted a single-Schechter function (Equation \ref{Eqn:Sch}), allowing $\alpha$, $M^*$, and $\Phi^*$ to freely vary, as well as with a fixed $\alpha = -1.3$ -- similar to methods adopted by a variety of studies \citep[e.g.,][]{Marchesini+09, Muzzin+13, Tomczak+14}. The fits are shown in the right panel of Figure \ref{fig:ODSMF}, and the best-fit parameters as well as their respective reduced-$\chi^2$ and BIC values are given in Table \ref{Tab:Sch}. As before, the covariances of the parameters in the Schechter fits are discussed in Appendix \ref{App:Schech}.

        We again performed a two-sample KS test to compare both the stellar mass distributions of Hyperion's outskirts and peaks to that of the field. We treated each MC realization separately, and plot the resulting distribution of $p$-values in the right plot of Figure \ref{fig:KS_Test}. When comparing mass distributions of Hyperion's outskirts and the field, we find 2/100 MC realizations with a $p$-value $<0.05$. On the other hand, when comparing the peaks and the field, we find 97/100 MC realizations have a $p$-value $<0.05$.

\begin{figure}
            \centering
            \includegraphics[width=1\linewidth]{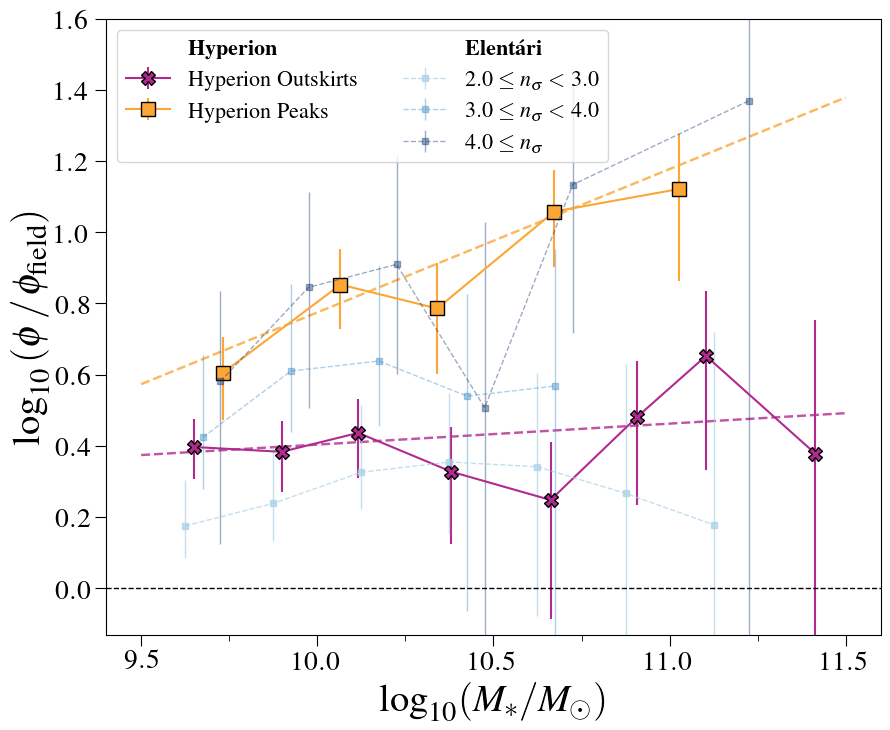}
            \caption{The SMFs of the Hyperion outskirts (magenta) and peaks (orange) after being normalized by the coeval field, and linear fits to both sets assuming $\log_{10}(\phi/\phi_{\rm field})=\beta\log_{10}(M_*)+\log_{10}(A)$. We additionally show the results from the Elent\`ari proto-supercluster at $z\sim 3.3$ \citep{Forrest+24}. In this framework, a flat line corresponds to an SMF in an overdense region with an identical shape to the coeval field, whereas a positive slope corresponds to an overdense region with a larger fraction of higher- to lower-mass galaxies. }
            \label{fig:SMFRatio}
        \end{figure}

        Finally, in Figure \ref{fig:SMFRatio} we plot the log-ratio of the SMFs in Hyperion and the field sample, along with similar measurements from the Elent\'ari protostructure at $z\sim3.3$ \citep{Forrest+24}, to highlight differences in their shapes. This effectively normalizes the SMFs such that one matching the field in shape will appear as a horizontal line with a vertical shift proportional to the overdensity, while an SMF with a relative excess of massive galaxies will show a positive slope.
        
        Empirically, trends appear approximately linear in log-log space. To quantify any trends, we fitted the log-ratios of both the outskirts and peaks with a power-law of the form $\phi/\phi_{\rm field} = A\cdot (M_*/M_\odot)^\beta$, where $A$ is a constant. This appears as a line with the form $\log_{10}(\phi/\phi_{\rm field})=\beta\log_{10}(M_*) + \log_{10}(A)$ when plotted in log-log space. The best-fit slopes, $\beta$, and reduced $\chi^2$ values are reported in Table \ref{Tab:PLaw}. A two-sided Student’s $t$-test performed during the linear regression analysis shows no significant correlation between stellar mass and the log-ratio in Hyperion’s outskirts relative to the field ($p = 0.38$), but a strong correlation in the peaks ($p = 0.02$). Despite the quality of the linear fits, one may expect the ratio of the SMFs to have a more complex shape given the relative complexity of the single-Schechter fits used to describe them. However, we show in Appendix \ref{App:SMFRatio} that a nearly linear fit (in log-log space) is expected in the mass range considered in this work.

        To assess the robustness of the above results to uncertainties inherent to photometric redshifts, we repeated the SMF analysis for only members of Hyperion with either a ground-based or \emph{HST}/grism spectroscopic sample. We find the difference between the SMFs of the overdensity peaks and the field is effectively unchanged, owing to the large fraction of galaxies in the peaks which have spectroscopic redshifts ($\sim 68/72$ in each MC iteration). The lack of difference between the shape of the SMF of Hyperion's outskirts and that of the field is also effectively unchanged, though the normalization of the SMF of the outskirts is lowered due to the smaller fraction of spectroscopic redshifts ($\sim 90/134$ in each MC iteration). Therefore, the results do not appear to be driven by the inclusion of galaxies with only photometric redshifts, or the scattering of such galaxies into and out of Hyperion in the Monte Carlo process.

        We also tested whether the SMF of overdensity peaks depends on large-scale substructure within Hyperion by subdividing the peak population according to redshift and right ascension. In particular, we find that the eastern ($\alpha\ge 150.2^\circ$) and western ($\alpha < 150.52^\circ$) portions of the lower-$z$ region ($z<2.52$) of Hyperion have similar SMFs. Because the most massive peak, Theia, is entirely contained within the western, lower-$z$ region (Table \ref{Tab:Peaks}), this agreement demonstrates that the observed SMF differences are not exclusive to the most massive and overdense region of Hyperion, but instead reflect a more general dependence on local overdensity.
        
        \begin{table}
            \caption{Parameters for the SMF ratios.}
            \label{Tab:PLaw}
            \centering
            
            \begin{tabular}{lcc}
                \hline\hline
                Sample & Power-law slope ($\beta$) & $\chi^2$/dof \\
                (1) & (2) & (3) \\
                \hline
                Hyperion Outskirts & $0.06 \pm 0.06$ & $0.54$  \\
                Hyperion Peaks  & $0.40 \pm 0.09$ & $0.37$  \\
                \hline
            \end{tabular}
            \tablefoot{The power-law fits for the SMF ratios ($\phi/\phi_{\rm field}$) shown in Figure \ref{fig:SMFRatio}. The columns are (1) the sample being fit; (2) the power-law slope of the fit, $\beta$, assuming $\log_{10}(\phi/\phi_{\rm field})=\beta\log_{10}(M_*)+\log_{10}(A)$; and (3) the reduced-$\chi^2$ value of each fit.}
        \end{table}

\section{Results and discussion} \label{Sect:Discussion}

    \subsection{Trends within the Hyperion SMF}

        Results from the analysis in Section \ref{sSect:Build_SMFs} suggest a consistent trend: the overdense peaks of Hyperion host galaxies with a different mass distribution than the field, having a higher ratio of high- to low-mass galaxies. A KS test indicates that 97\% of our MC realizations show a statistically significant difference between the SMFs of Hyperion’s peaks and the field (see right panel of Figure \ref{fig:KS_Test}). The nature of this difference is particularly apparent in the ratio of the two SMFs (Figure \ref{fig:SMFRatio}). The peaks have number densities of massive galaxies ($M_* \sim 10^{11}\,M_\odot$) that are $\sim 10\times$ higher than the field, compared to only $\sim 3\times$ higher at lower masses ($M_* \sim 10^{9.5}\,M_\odot$). Fitting a power law to this ratio, $\phi_{\mathrm{peak}}/\phi_{\mathrm{field}} \propto M_*^{\beta}$, we obtain $\beta = 0.40 \pm 0.09$, significantly different from $\beta=0$ ($t=4.5$, $p=0.02$), which would correspond to the SMFs having the same shape.

         This result is loosely corroborated by single-Schechter fits to the overdensity bins (right panel of Figure \ref{fig:ODSMF}; see also Table \ref{Tab:Sch}), though we caution against over-interpreting the best-fit values for Hyperion’s peaks. Degeneracies between $\Phi^*$, $M^*$, and $\alpha$ complicate direct comparisons between the parameters for fits of different regions. For the peaks in particular, $\Phi^*$ and $M^*$ are strongly degenerate (Figure \ref{Fig:Covars}), likely because the observed mass range lies entirely below the characteristic mass of the single-Schecther fit. In practice, this means the peak SMF is well described by a single power law, resulting in $\alpha$ being much more tightly constrained than either $\Phi^*$ or $M^*$ (Table \ref{Tab:Sch}; see also discussion in Appendix \ref{App:Schech}).

        Conversely, we find no evidence that the SMF in the outskirts of Hyperion is different from that of the field. A two-sample KS test shows no statistically significant difference between the SMFs of the outskirts and the field in 98\% of our MC realizations (right panel of Figure \ref{fig:KS_Test}). Furthermore, the Student's $t$-test suggests no correlation between the ratio of the two SMFs and stellar mass ($t=0.9$, $p=0.38$), which is apparent by the flat slope seen in Figure \ref{fig:SMFRatio}. Likewise, we find the total SMF of Hyperion also has a similar overall shape to that of the field (modulo the normalization factor) as seen in Figure \ref{fig:TotalSMF}. A two-sample KS test shows no statistically significant difference between the SMFs of Hyperion as a whole and the field in 77\% of our MC realizations, and single-Schechter fits yield $M^*$ and $\alpha$ values consistent with the field (Table \ref{Tab:Sch}).
        
        This is, perhaps, not unexpected given that the outskirts contain $\sim 2/3$ of the galaxies in Hyperion above $\log_{10}(M_*/M_\odot) \geq 9.5$ (on average $134_{-7}^{+10}$ versus $72_{-4}^{+4}$ in the peaks per MC realization). Therefore, total SMF is primarily dominated by the outskirts, with the peaks contributing only moderately having been slightly saturated after being normalized by larger volumes. This result highlights the importance of considering the protostructure within discrete overdensity bins rather than holistically, as environmental effects in overdense regions may be washed out in the total SMF.

        Qualitatively, the shape of the total SMF is interesting in that we see a slight dip at $\log_{10}(M_*/M_\odot)\sim 10.5$, and it is thus well fit by a double-Schechter function with a characteristic stellar mass at the point of the dip. However, the double-Schechter function has two additional degrees of freedom, and therefore has a slightly higher reduced-$\chi^2$ value and slightly lower BIC (Table \ref{Tab:Sch}). As such, it is somewhat ambiguous as to if the total SMF of Hyperion is more consistent with a single- or double-Schechter fit.

    \subsection{Broader context of the Hyperion SMF}

        The results of our SMF analysis suggest galaxies in the most overdense peaks of Hyperion are undergoing -- or have undergone -- some process which enhances the buildup of stellar mass relative to field galaxies. This trend of generally increasing stellar mass in the most overdense peaks is consistent with results of similar studies at intermediate redshifts \citep[$0.6<z<1.5$;][]{Tomczak+17, VanderBurg+20}, as well as other studies of protostructures at higher redshifts \citep[$z>2$;][]{Shimakawa+18a, Shimakawa+18b}. Additionally, in Figure \ref{fig:SMFRatio} we make a more direct comparison with results from \citet{Forrest+24}, who used a similar VMC method to study the Elent\'ari proto-supercluster at $z\sim 3.3$ in the COSMOS field. While a direct comparison between similar overdensities at different redshifts is difficult due to cosmic variance (see Figure \ref{Fig:OD_Dist}), the SMFs of both Hyperion and Elent\'ari suggest a similar trend of increasing stellar mass with increasing overdensity in high-redshift proto-superclusters.

        These trends likely reflect a broader shift in how environment influences galaxy evolution at high redshifts. It is well established that galaxies in clusters at low to intermediate redshifts ($z\lesssim 1.5$) tend to be quiescent and gas-poor, having built up their stellar mass at some earlier time \citep{Gomez+03, vonderLinden+10, Muzzin+12, Tomczak+19, Old+20}. However, recent studies have shown that protostructures at higher redshifts exhibit a fundamentally different relationship to the star formation rate (SFR) of their constituent galaxies. For instance, \citet{Lemaux+22} find a weak but statistically significant positive correlation between SFR and overdensity for $2\leq z \leq 5$, suggesting that early dense environments may in fact stimulate star formation, rather than suppress it. Likewise, a stacking analysis of infrared observations of protostructures revealed enhanced SFR densities (SFRDs) for protostructures in the redshift range $2\leq z\leq 4$ \citep{Kubo+19, Popescu+23}. In a more dedicated spectroscopic follow-up of the Taralay protocluster at $z\sim 4.57$, \citet{Staab+24} find an enhanced SFRD which they attribute to both a higher density of star-forming galaxies and an elevated SFR in said galaxies.

        The results of our analysis in Hyperion are consistent with -- and are perhaps a direct consequence of -- this positive correlation between overdensity and SFRs at higher redshifts. While it makes sense that elevated levels of star formation in high-redshift protostructures lead to more massive galaxies, the full process by which this might happen remains somewhat unclear. However, we offer a brief outline  using results of recent studies through which an enhancement of SFR in high-redshift overdense environments may result in the SMFs observed in Hyperion.

        By definition, overdense environments at high redshifts ($z\gtrsim4$) have a higher number density of galaxies than the coeval field. Dynamically, galaxies in these protostructure likely have slightly elevated velocity dispersions than the field, as a result of the larger gravitational potential of the overdense environment. It is therefore fair to assume that these galaxies experience more frequent galaxy-galaxy interactions than their counterparts in the field. At lower redshifts ($z \lesssim 1$), it has been shown this galaxy harassment can excite transient periods of star formation in dusty galaxies \citep{Moore+96, Moore+98}. Although in a recent study, \citet{Shah+22} show the typical SFR-enhancement in interacting galaxies is lower at higher redshifts ($1<z<3$) than at lower redshifts, even a modest increase in SFR produced by continual close encounters could result in a moderately elevated stellar mass in the $\sim 1.5$ Gyr between $z\sim 5$ and $z\sim 2.5$.

        Moreover, high-redshift overdense regions may not yet contain a fully heated intracluster medium \citep[ICM;][]{Overzier16, Shimakawa+18b}. Thus, any ambient diffuse gas which exists within the environment -- either as a result of the elevated gravitational potential of the environment or having been stripped during galaxy interactions -- could still be accreted onto the galaxies within the environment. Therefore, while these galaxies may sustain elevated SFRs, they may also benefit from more abundant gas reservoirs than field galaxies, resulting in sustained star formation and subsequent stellar mass enhancements.

        We now turn back to $z\sim 2.5$ to see if this picture holds for Hyperion and other protostructures. Within Hyperion, \citet{Giddings+26} find that the fraction of galaxies with close kinematic companions is nearly double that of the field, meaning galaxy-galaxy interactions appear to be elevated. Additionally, \citet{Gururajan+25} find a tentative positive correlation between SFR and overdensity within Hyperion, inline with the trend reported in \citet{Lemaux+22}, and which may be expected if there are more frequent close encounters. Interestingly, they also find that the gas fraction ($M_{\rm gas}/M_*$) of member galaxies within Hyperion tends to decrease with increasing overdensity, and appears uncorrelated in the most overdense peaks. Taken together with the increased SFRs, this tentatively implies an enhanced star formation efficiency ($\rm SFE\equiv SFR/M_{gas}$) within Hyperion, which is conceivably driven by close encounters of member galaxies. The flattening of gas-fraction trends in the most overdense peaks of Hyperion, combined with the decreasing star formation seen in other protostructure peaks at $z \lesssim 2.5$ \citep{Shimakawa+18a, Shimakawa+18b}, could signal the onset of a hot intracluster medium (ICM) in this redshift range. A recent study of the Spiderweb protocluster at $z\sim 2.16$ indeed detects signatures of an ICM via the thermal Sunyaev–Zeldovich effect \citep{SZ72, DiMascolo+23}. However, despite existing X-ray observations of Hyperion, it remains unclear whether an ICM has yet begun to develop \citep{Wang+16, Champagne+21}.

        On the other hand, sustained and elevated SFRs are not the only conceivable way to produce higher stellar masses in overdense environments. Observations suggest major mergers are another viable method through which galaxies at higher redshifts ($z\gtrsim2$) can experience substantial stellar mass buildup \citep{Tasca+14, Duncan+19, Ferreira+20}. In addition to adding much of the stellar masses of the two merging galaxies, it is thought that major mergers at this redshift can briefly enhance SFR and the prevelance of active galactic nuclei \citep[AGN; e.g.,][]{Hopkins+08}. Furthermore, through the comparison of a semi-empirical model and observed SMFs, \citet{Tomczak+17} demonstrate that galaxy-galaxy mergers play an important role in shaping the SMFs at intermediate redshifts ($0.6 < z < 1.3$). At these redshifts, intermediate overdensities appear dynamically favorable for galaxy mergers due to their moderate velocity dispersions \citep{Lin+10, Tomczak+17}.

        Correspondingly, the overdense peaks of higher-redshift structures, such as Hyperion, may be ideal locations for major galaxy mergers. As discussed previously, \citet{Giddings+26} find an elevated number of close kinematic pairs within Hyperion and estimate an average merging timescale of $\sim3-10$ Gyr. This timescale is consistent with the $\sim 3$ Gyr separating Hyperion at $z\sim2.5$ and the structures observed in \citet{Tomczak+17} at $z\sim 1$. Additionally, elevated fractions of major mergers in overdense protostructures are consistent with observations suggesting an increased AGN fraction at the redshift of Hyperion \citep{Shah+25}.

        Likely, the local environment of a galaxy in the high-redshift universe mediates enhanced stellar mass growth in more than one way. While the SMF of Hyperion presented in this paper provides compelling evidence that the environment does play a role at some level in promoting accelerated mass growth in the early universe, it is not enough to make a definitive argument as to what method dominates the process. Given the two mechanisms presented here -- enhanced SFRs via galaxy harassment and major mergers -- we might expect to find different quiescent fractions at different redshifts depending on which mechanism is primarily driving galaxy evolution. Whereas close encounters and harassment may lead to a more gradual quenching as the gas reservoir is heated and exhausted, major mergers may drive more rapid quenching through stellar feedback from starbursts and AGN feedback.

        Given the idea that overdense environments in the early universe may promote accelerated stellar mass growth, looking at quiescent fractions in and around these environments might provide insight into underlying physical mechanisms. \citet{Lin+15} find little correlation between local environment and quiescent fraction over $1.5 < z < 2.5$ in the COSMOS field, while a more recent study by \citet{Edward+24} finds a relative excess of low-mass ($\log_{10}(M_*/M_\odot)<10$) quiescent galaxies in protostructures at $2<z<2.5$, including Hyperion. However, this analysis relies solely on photometric classifications and does not explore the trend as a function of overdensity.  On the contrary, results from the ZFOURGE survey suggest the quiescent fraction is higher in more overdense environments up to $z\sim 2$ for galaxies with $\log_{10}(M_*/M_\odot)>9.5$ \citep{Kaw+17}, and up  to $z\sim 2.4$ for galaxies with $\log_{10}(M_*/M_\odot)>10.2$ \citep{Hartzenberg+23}. At higher redshifts still, \citet{Forrest+24b} find a quiescent fraction consistent with that of the field within the Elent\'ari proto-supercluster, though dense regions of the structure show evidence of enhanced quiescent fractions -- especially at higher stellar masses \citep{McConachie+22, McConachie+25}.

        These mixed findings highlight the complexity of environmental effects at high redshift and the need for more targeted studies. It is fair to assume that no one mechanism is the primary driver of this trend, or that the same mechanism dominates across different protostructures at similar redshifts. Ultimately, a more detailed analysis of the quiescent population within Hyperion (or lack thereof) as a function of overdensity may provide some insight as to the timescales of the physical mechanisms driving the enhanced stellar masses seen the SMFs presented here. Such work will be crucial to disentangling the complex interplay between environment, star formation, and mass growth in the high-redshift universe.

\section{Conclusion}\label{Sect:Conclusion}

    In this work, we construct the galaxy stellar mass function (SMF) in the Hyperion proto-supercluster at $z\sim 2.5$ by leveraging a combination of data from the COSMOS2020 catalog, various ground-based spectroscopic surveys, and a new targeted \emph{HST}-grism survey. To our knowledge, this is the most complete SMF in a single proto-supercluster at $z\gtrsim 2$ to date. The additional spectroscopy and improved photometric redshifts enable us to repopulate the existing Voronoi Monte Carlo (VMC) overdensity map of Hyperion with a much larger, more secure galaxy sample. This enriched map lets us compare, for the first time, the SMFs of Hyperion’s overdense peaks and its lower-density outskirts (Figure \ref{fig:ODSMF}).
    
    In doing so, we find two main results:
    \begin{enumerate}
        \item The overdense peaks in Hyperion contain a markedly larger fraction of high-mass galaxies than the coeval field (Figure \ref{fig:SMFRatio}). Specifically, we find that the peaks have number densities of massive galaxies ($M_* \sim 10^{11}\,\,M_\odot$) that are $\sim 10\times$ higher than the field, compared to only $\sim 3\times$ higher for lower-mass galaxies ($M_*\sim 10^{9.5}\,\,M_\odot$).

        \item Hyperion’s outskirts have an SMF shape indistinguishable from the field (Figure \ref{fig:SMFRatio}). Consequently, the total SMF of Hyperion also differs little from the field (Figure \ref{fig:TotalSMF}): the outskirts host nearly two-thirds of the galaxies in Hyperion, and their dominance dilutes the shallower SMF of the peaks when combined.
    \end{enumerate}

    Our first result aligns with recent work showing (i) elevated gas reservoirs and pair fractions in Hyperion, and (ii) a positive SFR–density correlation at $z\gtrsim 2$, opposite to the trend at lower redshift. Additionally, this high-mass enhancement echoes the trend seen in the Elent\'ari proto-supercluster at $z\sim3.3$. Taken together, these findings suggest that the densest regions of the early Universe {promote} rapid star formation and stellar-mass build-up rather than quenching it. A deeper census of the quiescent fraction and SFR density as functions of overdensity within Hyperion will be crucial for pinning down the timescales over which this accelerated growth operates.

    The second result highlights an important methodological point: studying high-redshift protostructures holistically may wash out environmental signals. In the case of the SMFs presented here, two effects combine: (i) the outskirts, whose SMF resembles the field, contain the majority of Hyperion’s galaxies, and (ii) the peak SMF is diluted when normalized over the larger protostructure volume. As a result, studies treating Hyperion as a whole, rather than dividing it into discrete overdensity bins, may incorrectly conclude there is no environmental impact on the SMF. This emphasizes the necessity of targeted spectroscopic surveys of high-redshift protostructures in order to expose environmental imprints that would otherwise be hidden in global averages.

\begin{acknowledgements}
We would like to thank the anonymous referee for their helpful comments.
DS is grateful for support through Prof.\ Colby Haggerty's NSF-CAREER grant AGS-2338131, which enabled the completion of this slight scientific detour.
BF acknowledges support from JWST-GO-02913.001-A. 
DCB is supported by an NSF Astronomy and Astrophysics Postdoctoral Fellowship under award AST-2303800. DCB is also supported by the UC Chancellor's Postdoctoral Fellowship. 
LG acknowledge the FONDECYT regular project number 1230591, the ANID - MILENIO - NCN2024\_112, and the ANID BASAL project FB210003 for financial support. 
Much of this work is based on photometry from the COSMOS2020 catalog, and we thus thank the COSMOS team for their efforts.
Some of the material presented in this paper is based upon work supported by the National Science Foundation under Grant No. 1908422.
This work is also based on observations collected at the European Southern Observatory under ESO programmes 175.A0839, 179.A-2005, and 185.A-0791.
This research is based on observations made with the NASA/ESA Hubble Space Telescope obtained from the Space Telescope Science Institute, which is operated by the Association of Universities for Research in Astronomy, Inc., under NASA contract NAS 5–26555.
Supported by the international Gemini Observatory, a program of NSF NOIRLab, which is managed by the Association of Universities for Research in Astronomy (AURA) under a cooperative agreement with the U.S. National Science Foundation, on behalf of the Gemini partnership of Argentina, Brazil, Canada, Chile, the Republic of Korea, and the United States of America.
This work is based on observations obtained with MegaPrime/MegaCam, a joint project of CFHT and CEA/IRFU, at the Canada-France-Hawaii Telescope (CFHT) which is operated by the National Research Council (NRC) of Canada, the Institut National des Science de l’Univers of the Centre National de la Recherche Scientifique (CNRS) of France, and the University of Hawaii.
This work is based in part on data products produced at Terapix available at the Canadian Astronomy Data Centre as part of the Canada-France-Hawaii Telescope Legacy Survey, a collaborative project of NRC and CNRS.
This work is based, in part, on observations made with the Spitzer Space Telescope, which is operated by the Jet Propulsion Laboratory, California Institute of Technology under a contract with NASA.
This research is based in part on data collected at the Subaru Telescope, which is operated by the National Astronomical Observatory of Japan.
Some of the data presented herein were obtained at Keck Observatory, which is a private 501(c)3 non-profit organization operated as a scientific partnership among the California Institute of Technology, the University of California, and the National Aeronautics and Space Administration. The Observatory was made possible by the generous financial support of the W. M. Keck Foundation.
The authors wish to recognize and acknowledge the very significant cultural role and reverence that the summit of Maunakea has always had within the Native Hawaiian community. We are most fortunate to have the opportunity to conduct observations from this mountain.
\\
\\
\emph{Software:} \texttt{Astropy} \citep{astropy}, \texttt{Matplotlib} \citep{MatPlotLib}, \texttt{NumPy} \citep{Numpy}, \texttt{SciPy} \citep{Scipy}, \texttt{statsmodels} \citep{statsmodels}.
\end{acknowledgements}

\bibliographystyle{aa}        
\bibliography{references}

\begin{appendix}

\section{The structure of Hyperion}\label{App:HyperionStructure}

\begin{figure*}
            \centering
            \includegraphics[width=\linewidth]{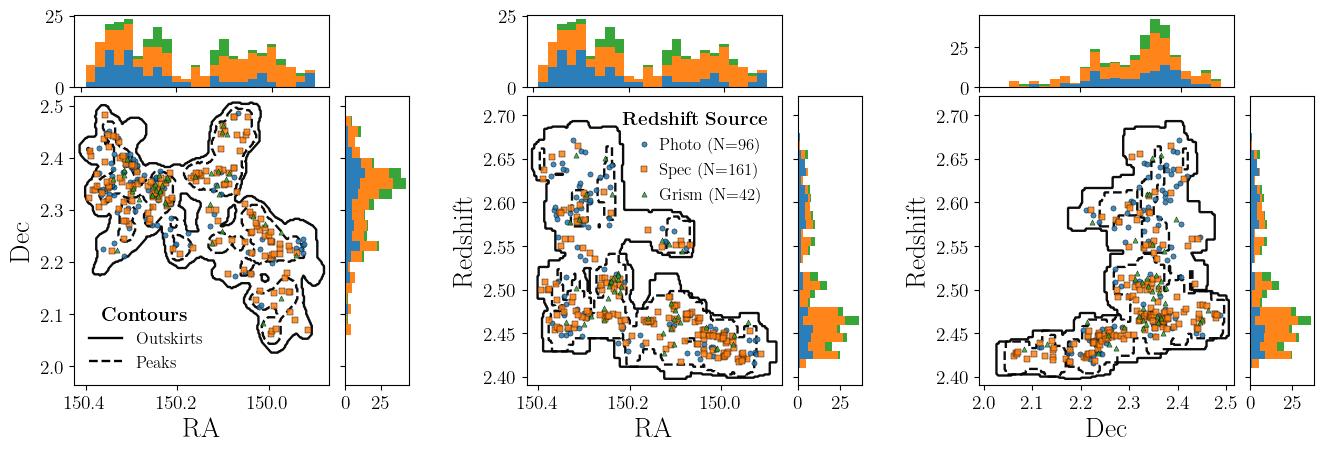}
            \caption{The distribution of Hyperion member galaxies, projected in the (left) R.A.-Dec., (middle) R.A.-$z$, and (right) Dec.-$z$ planes. In each panel, the solid and dashed contours correspond to Hyperion's outskirts and peaks, respectively, as defined by the VMC overdensity map. Additionally, the locations of the member galaxies are plotted prior to any Monte Carlo redshift sampling, and their marks are distinguished according to the redshift source: COSMOS2020 photometry (blue circles), ground-based spectroscopy (orange squares), or \emph{HST}/grism spectroscopy (green triangles). For galaxies with both a ground-based spectroscopic redshift and a  \emph{HST}/grism redshift, the redshift with the highest quality according to the quality flags discussed in Section \ref{Sect:Data} is chosen. The histograms along each axis show the distribution of member galaxies in the corresponding projected dimension, color-coded by redshift source.
            }
            \label{Fig:HyperionMems}
        \end{figure*}

    In this section, we provide additional details on the structure of Hyperion. Figure \ref{Fig:HyperionMems} shows 2D projections of Hyperion in the R.A.-Dec., R.A.-$z$, and Dec.-$z$ planes. Also plotted are the member galaxies of Hyperion (before any Monte Carlo sampling of redshifts) and their redshift sources. 

    \subsection{Peaks of Hyperion}

        As described in Section \ref{Sect:GalEnvs}, we define a peak as a set of contiguous voxels with $n_\sigma \geq 4.0$ (Equation \ref{Eqn:PolyFit}). Multiple overdensity peaks are identified within Hyperion, spanning a range of masses and volumes, but the SMF of the peaks is dominated by the most massive structures that contain the majority of galaxies.
        
        Table \ref{Tab:Peaks} lists the properties of all peaks that contain, on average, at least one galaxy per MC iteration. Peak masses are the sums of the constituent voxel masses, which are given by Equation \ref{Eqn:PeakMass}. The peak volumes are the sums of their voxel volumes. The average number of galaxies reported corresponds to those with $\log_{10}(M_*/M_\odot) \geq 9.5$ across 100 MC realizations. On average, $\sim 72$ galaxies contribute to the peak SMF, and the seven most massive peaks listed here contain $\sim 98\%$ of them.
        
        \begin{table*}
            \caption{Overdensity peaks in Hyperion.}
            \label{Tab:Peaks}
            \centering
            \begin{tabular}{lcccccc}
                \hline\hline
                ID &  $\rm R.A.$ & $\rm Dec.$ & $z$ & $M_{\rm tot}$  & Volume & $\rm\langle N_{gal}\rangle$ \\
                & (deg) & (deg) & & ($10^{14}\,\,M_\odot$) & ($\rm cMpc^{3}$) &  \\
                (1) & (2) & (3) & (4) & (5) & (6) & (7) \\
                \hline
                1 & $150.097$ & $2.388$  & $2.467$  & $2.02$ & $2791$ & $21.9$ \\
                2 & $150.259$ & $2.344$  & $2.464$  & $0.92$ & $1344$ & $13.6$ \\
                3 & $150.005$ & $2.248$  & $2.442$  & $0.88$ & $1312$ & $15.6$ \\
                4 & $149.985$ & $2.122$  & $2.427$  & $0.48$ & $724$ & $3.2$ \\
                5 & $150.095$ & $2.352$  & $2.554$  & $0.22$ & $316$ & $3.6$ \\
                6 & $150.238$ & $2.335$  & $2.505$  & $0.43$ & $658$ & $11.1$ \\
                7 & $150.104$ & $2.238$  & $2.436$  & $0.13$ & $202$ & $1.5$ \\
                \hline
            \end{tabular}
            \tablefoot{The various overdensity peaks (as defined in Section \ref{sSect:IDing_Protos}) within Hyperion with at least one galaxy in each MC iteration on average. The columns correspond to  (1) peak number, central; (2) right ascension; (3) declination;  (4) redshift; (5) total mass; (6) total volume (not corrected for elongation); and (7) average number of galaxies in each MC iteration.}
        \end{table*}

    \subsection{High-redshift offshoot}

        The updated VMC maps used here extend Hyperion to higher redshifts than in \citet{Cucciati+18}. This could represent a genuine extension of the structure, newly connected through improved spectroscopic completeness. If so, it is unclear whether galaxies in this region share the same evolutionary pathways as those in the main body of Hyperion, or how long the two portions will evolve together.
        
\begin{figure}
            \centering
            \includegraphics[width=\linewidth]{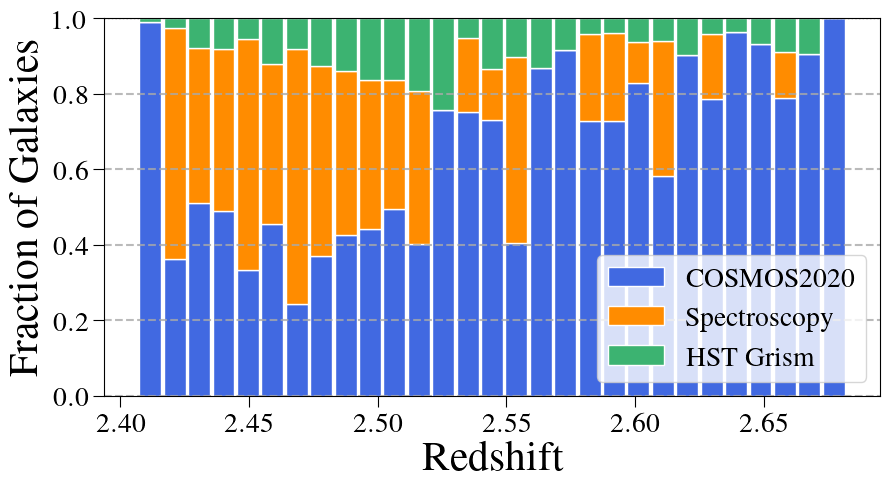}
            \caption{We show, as a function of redshift, the fraction of galaxies within Hyperion across all 100 MC iterations with redshifts determined from photometry (blue), ground-based spectroscopy (orange), and \emph{HST}-grism spectroscopy (green). The lower-redshift portion of Hyperion is dominated by spectroscopic redshifts, whereas the higher-redshift portion is dominated by photometric redshifts. }
            \label{Fig:ot_frac}
        \end{figure}

        A more likely explanation, however, is that this apparent offshoot is the result of redshift-space smearing in the VMC maps. As shown in Figure \ref{Fig:ot_frac}, the fraction of galaxies assigned via photometric redshifts rises from $\sim 40\%$ in the main body ($z \lesssim 2.52$) to $\sim 80\%$ at higher redshift. These galaxies may or may not truly belong to Hyperion, and some likely have broad redshift probability distributions.
        
\begin{figure}
            \centering
            \includegraphics[width=\linewidth]{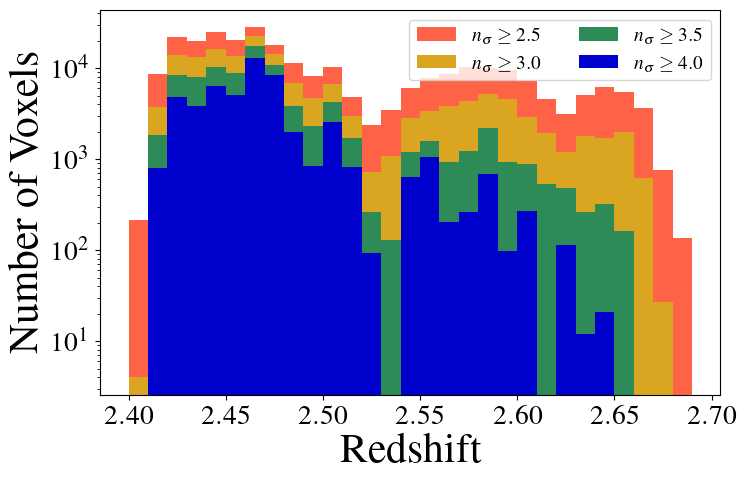}
            \caption{We show the distribution of voxels within Hyperion for four different overdensity thresholds: $n_\sigma \geq 2.5$ (orange), $n_\sigma \geq 3.0$ (yellow), $n_\sigma \geq 3.5$ (green),  $n_\sigma \geq 4.0$ (blue). The low- and high-redshift portions are connected at $z\sim 2.53$ until an overdensity cut of $n_\sigma \geq 4.0$.}
            \label{Fig:vox_dist}
        \end{figure}

        To assess the robustness of this feature, Figure \ref{Fig:vox_dist} shows the overdensity distribution of voxels as a function of redshift. Excluding the high-$z$ extension would require either (i) raising the overdensity threshold for outskirts to $3.5 \leq n_\sigma < 4.0$, or (ii) retaining $n_\sigma \geq 2.5$ but imposing an artificial cut at $z\sim2.52$. Either option seems disingenuous in an attempt to objectively map the structure of Hyperion from the VMC maps available. We therefore retain the extension in our maps, while cautioning that it is likely a mapping artifact rather than a distinct physical component.
        
        Importantly, the inclusion of this feature does not affect the main result of this work: that the peaks of Hyperion show an enhanced abundance of massive galaxies relative to the field. Of the seven peaks listed in Table \ref{Tab:Peaks}, only one lies in the high-redshift offshoot, and it contributes just $\sim 5\%$ of the galaxies in the peak SMF on average. Moreover, we find that including the offshoot does not alter the similarity between Hyperion’s outskirts and the field, indicating that this result is not simply driven by field galaxies scattering into Hyperion.
        
\section{Schechter fits}\label{App:Schech}

    As with most studies of the stellar mass function (SMF), we model our measurements with single-Schechter functions (Equation \ref{Eqn:Sch}). The best-fit parameters minimizing $\chi^2$ are reported in Table \ref{Tab:Sch}. While these fits are straightforward to compute, their interpretation is complicated by parameter degeneracies, particularly for less well-constrained populations such as the peaks of Hyperion. In this Appendix, we illustrate these degeneracies and assess what conclusions can still be drawn.
    
\begin{figure*}
    \centering
    \includegraphics[width=\linewidth]{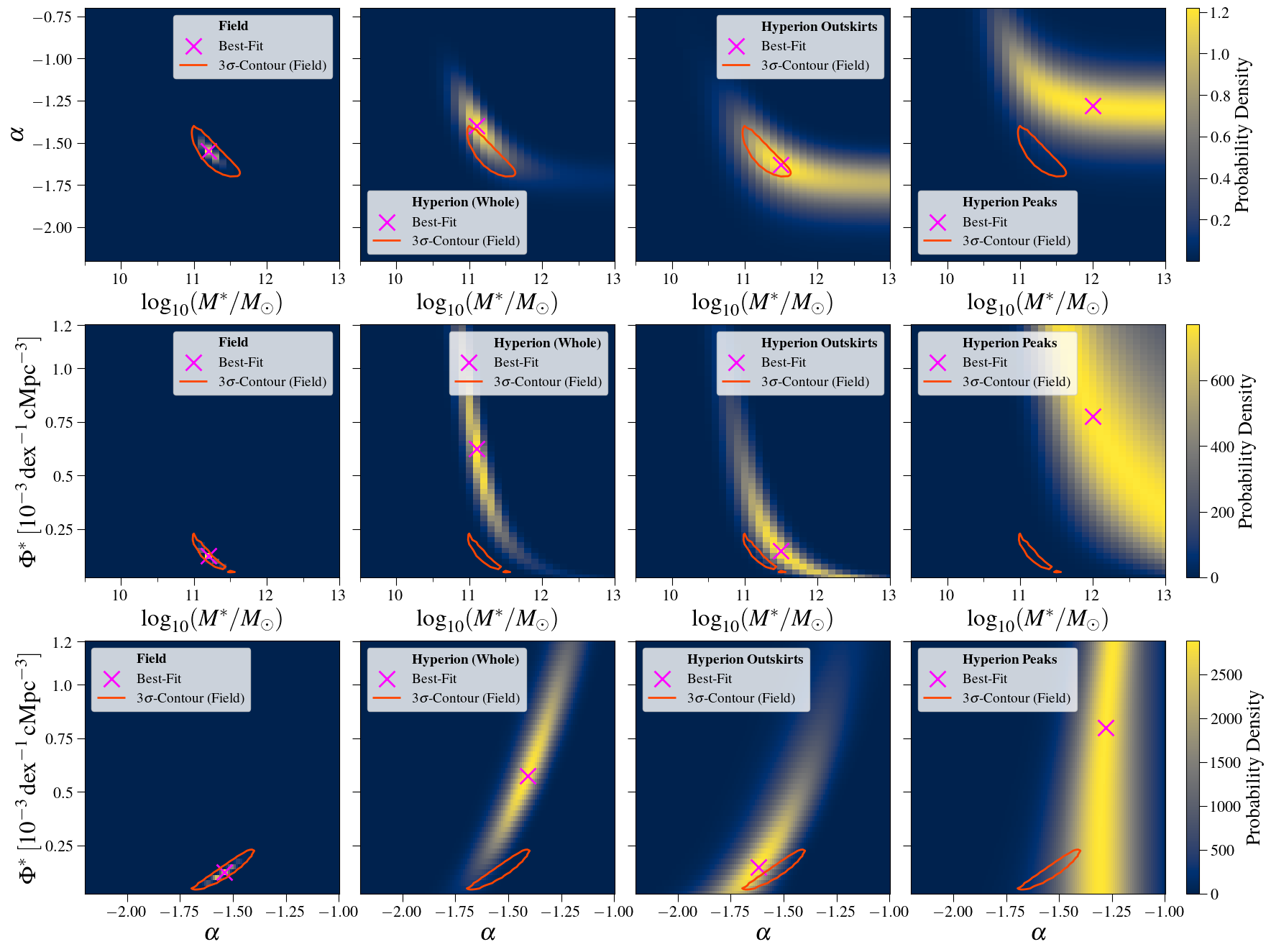}
    \caption{To explore the degeneracies of the parameters in the single-Schechter functions, we plot the parameter spaces of the fits to the SMFs of (Columns 1-4) the field, Hyperion as a whole, Hyperion's outskirts, and Hyperion's peaks. In each row, we fit over a grid of two of the three parameters in the single-Schechter function, and marginalize over the remaining parameter to minimize the $\chi^2$ value. The parameter spaces explored are (row 1) $\log_{10}M^*-\alpha$, (row 2) $\log_{10}M^*-\Phi^*$, and (row 3) $\alpha-\Phi^*$. For each subplot, we  plot the point with the minimum $\chi^2$ value in the given parameter space (magenta X). In each row, we additionally plot a $3\sigma$ contour from the field, corresponding to $\Delta\chi^2\simeq 11.8$, to make comparisons between the likelihood surfaces from the field and the other populations in each column easier.}
    \label{Fig:Covars}
\end{figure*}

   To visualize these degeneracies, we map the likelihood surface across the Schechter parameter space for four populations: the field, Hyperion as a whole, its outskirts, and its peaks. For each population, we fix two parameters over a grid of plausible values and fit the third by minimizing $\chi^2$. The top row of Figure \ref{Fig:Covars} shows an example: $\Phi^*$ is fit freely at each grid point in $(\log_{10}M^*,\alpha)$. Each grid point thus has an associated $\chi^2_i$, and the corresponding likelihood is $\mathcal{L}_i \propto \exp[-(\chi^2_i-\chi^2_{\rm min})/2]$. Normalizing by the sum of all likelihoods yields a probability distribution function (PDF). Confidence regions are then defined by contours of $\Delta\chi^2 = \chi^2_i - \chi^2_{\rm min}$ appropriate for two free parameters, with the $3\sigma$ contour corresponding to $\Delta\chi^2 \simeq 11.8$. We emphasize that these confidence intervals are evaluated in the two-dimensional subspace of fixed parameters; the third parameter is always fit freely and is therefore marginalized over rather than explicitly constrained \citep[see][]{Avni76}.

    For the field population (first column of Figure \ref{Fig:Covars}), the likelihood surface is compact and well-behaved: the three parameters are simultaneously constrained, and the $3\sigma$ contours are tight. The resulting values agree with previous measurements in similar redshift ranges \citep[e.g.,][]{Weaver+23}, providing confidence in both the data and the fitting procedure. In the second column, the parameter space for Hyperion as a whole mirrors the conclusions drawn in the main text. Both of the bottom panels highlight its elevated normalization relative to the field—a direct consequence of its overdense nature. The top panel shows that Hyperion has a marginally shallower power-law slope and a characteristic mass similar to the field, with the $(\log_{10}M^*,\alpha)$ likelihood surface nearly co-spatial with that of the field. The outskirts of Hyperion (third column) likewise exhibit a slope comparable to the field but a slightly higher normalization, consistent with the trend observed in Figure \ref{fig:SMFRatio}.

    The peaks of Hyperion (fourth column) stand out in that some parameters are effectively unconstrained. This likely reflects that the observed SMF is better approximated by a pure power-law: the “knee” of the Schechter function is not detected, leaving $M^*$ unresolved. As a result, $\alpha$ is relatively well constrained, while $M^*$ and $\Phi^*$ remain strongly degenerate. This degeneracy is most apparent in the bottom panel, where the likelihood surface is nearly vertical around the best-fit $\alpha$. The large uncertainties on $M^*$ and $\Phi^*$ in Table \ref{Tab:Sch} echo this behavior, with the normalization errors in particular inflated by the effective freedom of $M^*$. Thus, while the formal uncertainties on $M^*$ and $\Phi^*$ for Hyperion’s peaks are large, they arise precisely because the SMF in these regions differs from that of the field: the low-mass slope is shallower, the characteristic mass is shifted to higher values and not observed, and the distribution is therefore better described by a power law.

\begin{table*}[!h]
\caption{Schechter function fits for various SMFs.}
\label{Tab:Sch}
\centering
\begin{tabular}{lcccccccc}
\hline\hline

Fit & $\log_{10} (M^*/M_\odot)$ & $\alpha_1$ & $\Phi_1^*$\tablefootmark{$\dagger$} & $\alpha_2$ & $\Phi_2^*$\tablefootmark{$\dagger$} & $\chi^2$/dof & BIC & Figure \\
(1) & (2) & (3) & (4) & (5) & (6) & (7) & (8) & (9) \\

\hline

Single-Schechter  & $11.13 \pm 0.22 $ & $-1.41 \pm 0.14 $ & $0.58 \pm 0.34 $ &  --- & --- & $0.807$ & $10.28$ &  \ref{fig:TotalSMF} \\

Double-Schechter & $10.43 \pm 0.18  $ & $-1.28 \pm 0.25  $ & $1.67 \pm 0.95  $ & $1.77 \pm 1.27  $ & $0.48 \pm 0.58  $ & $0.505$ & $11.91$ & \ref{fig:TotalSMF} \\

Field\tablefootmark{$\dagger\dagger$} & $10.97^{+0.06}_{-0.07}$ & $-1.46$             & $0.24^{+0.03}_{-0.02}$ & --- & ---  & 27.413 & --- & \ref{fig:TotalSMF} \\

\hline

Field & $11.21 \pm 0.10 $ & $-1.55 \pm 0.05 $ & $0.12 \pm 0.03 $ &  --- & --- & $1.305$ & $12.76$ &  \ref{fig:ODSMF} \\
      & $10.89 \pm 0.08 $ &  $-1.3$ & $0.32 \pm 0.04 $ &  --- & --- & $6.271$ & $41.79$ &   \ref{fig:ODSMF} \\

Hyperion Outskirts & $11.49 \pm 0.36 $ & $-1.63 \pm 0.10 $ & $0.15 \pm 0.13 $ &  --- & --- & $0.258$ & $7.53$ & \ref{fig:ODSMF} \\
        & $10.94 \pm 0.14 $ &  $-1.3$ & $0.73 \pm 0.14 $ &  --- & --- & $0.671$ & $8.18$ &  \ref{fig:ODSMF} \\

Hyperion Peaks  & $12.00 \pm 4.52 $ & $-1.28 \pm 0.29 $ & $0.78 \pm 3.36 $ &  --- & --- & $0.784$ & $6.40$ &  \ref{fig:ODSMF}\\
            & $12.37 \pm 3.68 $ &  $-1.3$ & $0.55 \pm 1.45 $ &  --- & --- & $0.523$ & $4.79$ &  \ref{fig:ODSMF}  \\

\hline
\end{tabular}
\tablefoot{ The best-fit Schechter function fits for various SMFs throughout the paper. The parameters for single- and double-Schechter functions correspond to equations \ref{Eqn:Sch} and \ref{Eqn:DSch}, respectively. The clumns correspond to (1) what is being fit, (2) the characteristic mass, (3) the power-law slope of single-Schechter fit, (4) normalization of single-Schechter fit, (5) power-law slope of double-Schechter fit, (6) normalization for double-Schechter fit, (7) reduced-$\chi^2$ value of the fit, (8) the Bayesian information criterion of the fit (via Equation \ref{Eqn:BIC}), and (9) the figure the fit can be seen in.\\
\tablefoottext{$\dagger$}{Normalizations are in units of $\rm 10^{-3}\,\,dex^{-1}\,\,Mpc^{-3}$.}\\
\tablefoottext{$\dagger\dagger$}{The field fit in Figure~\ref{fig:TotalSMF} is taken from the fit for $2.5 < z \leq 3.0$ in \citet{Weaver+23}.}
}
\end{table*}

\section{SMF ratio fits}\label{App:SMFRatio}

    In order to highlight the difference between the SMF of the peaks in Hyperion and the field, we analyze the log-ratio of the two (see Figure \ref{fig:SMFRatio}). As part of the analysis, we use a simplistic model in which $\log_{10}(\phi/\phi_{\rm field}) = \beta\log_{10}(M_*/M_\odot) + \log_{10}(A)$. However, given the relative complexity of the Schechter function, it is not immediately obvious why the ratio should take this form. To explore this more, we briefly derive an analytical expression for the ratio, $\log_{10}(n_{\rm gal}^H/n_{\rm gal}^f)$, in terms of the parameters of the two single-Schechter functions. Given Equation \ref{Eqn:Sch}, and defining $\mu_H\equiv \log_{10}(M_*/M^*_H)$ and $\mu_f\equiv \log_{10}(M_*/M^*_f)$, it follows that

    \begin{align*}
    \begin{split}
        \frac{n_{\rm gal}^H}{n_{\rm gal}^f} = \frac{\Phi^*_H}{\Phi^*_f}\exp\left(10^{-\mu_H} + 10^{-\mu_f}\right) \\
        \qquad \qquad\times 10^{ \left[(\alpha_H+1)\mu_H - (\alpha_f+1)\mu_f\right] }.
    \end{split}
    \end{align*}
    
    Upon simplifying this expression, one arrives at an expression for the log-ratio of the SMFs as a function of stellar mass, $M_*$, given as
    \begin{align*} 
        \log_{10}\left(\frac{n_{\rm gal}^H}{n_{\rm gal}^f}\right) &= \left[\alpha_H - \alpha_f\right]\cdot \log_{10}\left(\frac{M_*}{M_\odot}\right)\\ 
         &+ \log_{10}(e)\left[10^{\mu_f} - 10^{\mu_H}\right] + \mathcal{M}_0,
    \end{align*}
    
    where, for brevity, we have introduced the following constant:
    \begin{equation*}
    \begin{split}
        \mathcal{M}_0 \equiv (\alpha_f+1)\cdot\log_{10}\left(\frac{M^*_f}{M_\odot}\right) &- (\alpha_H+1)\cdot\log_{10}\left(\frac{M^*_H}{M_\odot}\right) \\
        & \quad+ \log_{10}\left(\frac{\Phi^*_H}{\Phi^*_f}\right).
        \end{split}
    \end{equation*}

    The first term in this expression is primarily what is reflected in Figure \ref{fig:SMFRatio} -- the ratio of the SMFs is primarily a power law with an exponent equal to the difference of the low-mass slopes of the two SMFs. However, we also see that the ratio really only follows this simple form if $\mu_f\approx \mu_H$, in which case the second term is nearly zero. Logically, this is satisfied unless the two SMFs have dramatically different characteristic masses. In that case, for masses larger than the smaller of the two characteristic masses, one SMF is exponentially cutoff while the other is a power law, and the ratio blows up as a result.

\begin{figure}
        \centering
        \includegraphics[width=\linewidth]{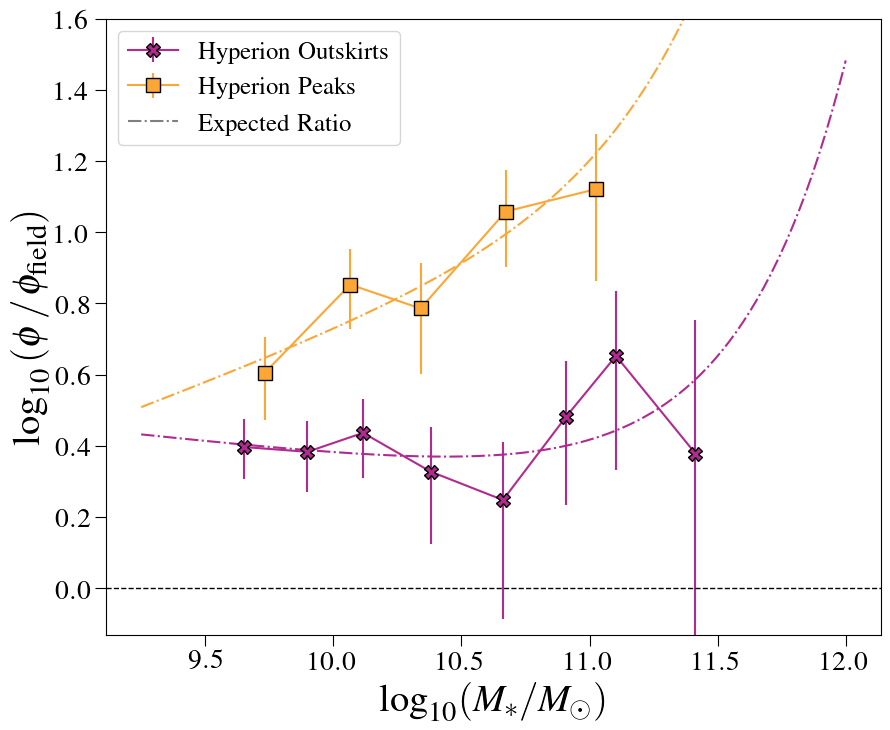}
        \caption{The SMFs of the Hyperion outskirts (magenta) and peaks (orange) after being normalized by the coeval field. The points are identical to those shown in Figure \ref{fig:SMFRatio}. The lines are the expected forms of the ratios using the best-fit parameters for the single-Schechter functions. For most the stellar mass range considered in this work, the ratio is expected to be nearly linear in log-log space. }
        \label{Fig:RatioTheory}
    \end{figure}

    In practice, the log-ratios of Hyperion’s peaks and outskirts to the field are close to linear. Figure \ref{Fig:RatioTheory} shows the expected forms using the single-Schechter parameters from Table \ref{Tab:Sch}. For the outskirts, the characteristic mass is similar to that of the field, so the ratio is approximately a power-law with a slope $\Delta\alpha \simeq -0.08$, yielding an almost flat relation. On the other hand, the SMF of the peaks has a shallower power law than the field, and thus the log-ratio has a steadily rising slope of $\Delta\alpha \simeq +0.27$. In principle, an exponential divergence should appear once the field SMF turns over at $M^*_f$, but this occurs near the upper limit of our mass range. Thus, across the observed bins the ratio is well described by a simple power law. We note that the slopes quoted here are slightly different from those obtained via regression fits to the binned ratios in the main text (Table \ref{Tab:PLaw}). This modest discrepancy reflects the finite mass range of the data and the onset of exponential suppression in the field SMF, but both approaches yield consistent qualitative trends.

\end{appendix}

\end{document}